\documentclass[3p,final]{elsarticle}

\usepackage[utf8]{inputenc}
\usepackage[english]{babel}

\usepackage{graphicx}

\usepackage{hyperref}

\usepackage{amsmath}
\usepackage{amssymb}
\usepackage{mathdots}
\usepackage{latexsym}

\biboptions{numbers,sort&compress}

\renewcommand{\vec}[1]{\ensuremath{\mathbf{#1}}}
\newcommand{\pd}[2]{\frac{\partial{} #1}{\partial{} #2}}
\providecommand{\vr}{\vec{r}}
\renewcommand{\vr}{\vec{r}}
\newcommand{\vR}{\vec{R}}
\newcommand{\rd}{{\rm d}}
\newcommand{\ii}{\mathrm{i}}
\newcommand{\ee}{\mathrm{e}}
\newcommand{\vnabla}{\boldsymbol{\nabla}}
\newcommand{\matrixel}[3]{\left< #1 \vphantom{#2#3} \right| #2 \left| #3 \vphantom{#1#2} \right>}

\newcommand{\ie}{\textit{i.e.}}
\newcommand{\eg}{\textit{e.g.}}

\newcommand{\sumall}{\sum_{\substack{i\,k\,n\\i'k'n'}}}
\newcommand{\intall}{\int_{\Omega}\rd^3r\,}

\DeclareMathOperator{\sgn}{sgn}

\begin{document}

\title{Scalable Atomistic Simulations of Quantum Electron Transport\\using Empirical Pseudopotentials}

\author{Maarten~L.~Van de Put\corref{cor}}
\ead{vandeput.maarten@gmail.com}
\cortext[cor]{Corresponding author}
\author{Massimo~V.~Fischetti}
\author{William~G.~Vandenberghe}

\address{%
Department of Materials Science and Engineering,
The University of Texas at Dallas\\%
800 W. Campbell Rd., Richardson, TX 75080, USA}

\date{\today}

\begin{abstract}
The simulation of charge transport in ultra-scaled electronic devices requires the knowledge of the atomic configuration and the associated potential.
Such ``atomistic'' device simulation is most commonly handled using a tight-binding approach based on a basis-set of localized orbitals.
Here, in contrast to this widely-used tight-binding approach, we formulate the problem using a highly accurate plane-wave representation of the atomic (pseudo)-potentials.
We develop a new approach that separately deals with the intrinsic Hamiltonian, containing the potential due to the atomic configuration, and the extrinsic Hamiltonian, related to the external potential.
We realize efficient performance by implementing a finite-element like partition-of-unity approach combining linear shape functions with Bloch-wave enhancement functions.
We match the performance of previous tight-binding approaches, while retaining the benefits of a plane wave based model.
We present the details of our model and its implementation in a full-fledged self-consistent ballistic quantum transport solver.
We demonstrate our implementation by simulating the electronic transport and device characteristics of a graphene nanoribbon transistor containing more than 2000 atoms.
We analyze the accuracy, numerical efficiency and scalability of our approach.
We are able to speed up calculations by a factor of 100 compared to previous methods based on plane waves and envelope functions.
Furthermore, our reduced basis-set results in a significant reduction of the required memory budget, which enables devices with thousands of atoms to be simulated on a personal computer.
\end{abstract}

\maketitle

\section{Introduction}

The numerical study of electron transport in solid-state transistors provides an important contribution to the improvement of future electronic devices.
To keep ahead of technological progress, the methods used to predict electron transport behaviour have shifted from simplified quasi-classical methods to advanced quantum mechanical descriptions.
Historically, this evolution has been driven by the continual reduction of the length-scales to dimensions at which the classical limit is no longer appropriate.
More recently, novel materials have been considered to improve the performance of electronic devices.
For example, atomically thin monolayers, such as graphene,~\cite{Schwierz2010} phosphorene,~\cite{Gaddemane2018} and transition-metal dichalcogenides,~\cite{Giacometti2011,Laturia2018} and their ribbons,~\cite{Llinas2017,Fang2017,Fang2016,Fischetti2011} are being actively investigated as possible replacements of silicon as the channel material in field-effect transistors.
These materials have caused an additional shift from transport models based on bulk-material properties towards the comprehensive modeling of the atomic structure of the material.
Whereas an atomistic description of quantum electron transport is widely applicable to different materials and device structures, atomistic resolution comes at a significant computational expense.

The atomistic calculation of the electronic structure starts by selecting an appropriate set of basis functions to discretize the problem.
Two popular approaches, each at one end of the spectrum, are the Linear Combination of Atomic Orbitals (LCAO), which is closely related to the picture of chemical bonding, and plane-wave based methods which form a natural basis for the physics of periodic crystals.
The most commonly used approximation of LCAO is the tight-binding (TB) approximation in which the interaction of the localized orbitals is short range, often only nearest neighbor (NN) orbitals being taken to overlap.~\cite{Fonseca2013,Garcia2002}
However, as remarked by Slater~\cite{Slater1937}, in the interstitial region, away from the ionic cores, the wavefunction in a crystalline solid is plane-wave like.
Due to the lack of non-bound states (\ie, ``scattering'' or ``traveling'' wavefunctions) in the tight-binding basis, its accuracy is limited when describing higher energy valence and conduction states where electrons are located in the interstitial region.
On the other hand, the plane-wave basis is a complete set whose accuracy can be carefully controlled by changing its truncation through a cutoff of the kinetic energy.
However, to describe the core-states accurately, a high energy-cutoff is needed to obtain a sufficiently fine spatial resolution in the region close to the ionic cores, a region that is more easily described by localized orbitals.
For this reason, all-electron calculations often feature hybrid methods, using plane waves to describe the interstitial regions, augmented with a localized basis to capture the core states.~\cite{Weinert2009,Blochl1994}

For the purposes of electron transport, we are interested in an accurate representation of the highest valence and lowest conduction states, which are, as discussed before, best captured by a plane-wave basis.
However, plane waves are, by definition, not localized and interactions between all plane waves need to be considered; resulting in dense linear algebra formulations that have a high computational burden compared to the sparse linear algebra that results from tight-binding methods.
For this reason the tight-binding approach is currently the most commonly used method to study quantum electron transport; using either a predefined set of orbitals with empirical parameters, \eg, the well known $\mathrm{sp}^3\mathrm{d}^5\mathrm{s}^*$ set, or using maximally localized Wannier functions to calculate the local orbitals from first-principles.~\cite{Fonseca2013,Bruck2017,Luisier2014,Maassen2013}
Commercial tight-binding transport simulators have already been developed to complement Technology Computer Aided Design (TCAD) in the semiconductor industry~\cite{QuantumATK}.
More limited investigations of plane-wave based transport has been undertaken academically, both based on \textit{ab-initio} pseudopotentials~\cite{Garcia-Lekue2015,Choi1999} and empirical pseudopotentials~\cite{Fang2017,Fang2016}. 
In addition to the high accuracy of these plane-wave methods, they allow us to probe locally or disturb the interstitial region with impurities and local fields, for example.
However, plane-wave methods have been applied only to relatively small atomic structures (up to 1000s of atoms) due to their computational burden, and even for these small systems they require expensive high-performance computing infrastructure.

In this paper, we develop a method that combines the computational benefits of the tight-binding approach, while maintaining the versatility and accuracy of plane-wave methods to represent the real-space wavefunctions throughout the atomic structure.
To achieve this goal, we turn to the Bloch waves of the crystal as an alternative basis to plane waves and tight-binding orbitals.
The benefits of using Bloch waves have been described for non-atomistic models in the context of the linear combination of bulk bands (LCBB) method~\cite{Wang1997,Jiang2012,Jiang2011,Esseni2005} and a recently developed empirical pseudopotential method for confined nanostructures~\cite{Pala2018}.
In contrast to these methods, our method relies on an expansion on the Bloch-waves of the atomic structure.
This enables the treatment of atomistic nano-structures that do not have a bulk crystal counterpart or whose electronic structure is dissimilar to the bulk material, \eg, carbon nanotubes, graphene nanoribbons, and extremely small silicon nanowires.
In addition, the atomistic nature of our method provides access to the atomic positions which enables the study of lattice defects and impurities in a straightforward way.

We focus on transport through nanostructures featuring one-dimensional transport, \ie, where the carriers are sufficiently confined such that they have only one degree of freedom.
To describe the electronic structure of these nanostructures, we adopt the atomistic empirical pseudopotential approximation~\cite{VandePut2016,Fang2017,Fang2016,Fischetti2011,Kim2011}.
Note that we make the distinction between bulk and atomistic empirical pseudopotential methods.
In the bulk empirical pseudopotential method, it is sufficient to know the values of the pseudopotential only at discrete reciprocal lattice vectors (form-factors).
In our method, which we call the atomistic empirical pseudopotential method, the pseudopotential $V(\vec{q})$ is given as a function of a wave vector $\vec{q}$ in reciprocal space, yielding a more general method.
Care must still be taken when transferring the pseudopotential from one system to another since one cannot expect, \textit{a-priori}, that different atomic configurations can be described by a non self-consistent pseudopotential.
However, there are known cases, such as the set of pseudopotentials for carbon nanostructures, introduced by Kurokawa~\cite{Kurokawa2000}, that show unexpected good performance for a wide range of atomic structures, including the graphene nanoribbons, we study as an example of a one-dimensional nanostructure in this work.

Our paper is structured as follows.
In Section~\ref{s:theory}, we discuss the models for the atomic and electronic structure and develop the theory of our Bloch-wave basis. 
Section~\ref{s:transport} details the calculation of the electronic properties in an open system with contacts.
In Section~\ref{s:self-consistent}, we explain the self-consistent procedure, coupling the electrostatics with the electron density in the system.
Section~\ref{s:results} shows the application of our method to an armchair graphene-nanoribbon transistor, including verification of the accuracy and computational efficiency of our approach.
Finally, we conclude in Section~\ref{s:conclusions}

\section{Theoretical model}%
\label{s:theory}

\subsection{Model Hamiltonian}

To model electron transport in nanoscaled devices, two length-scales should be considered: (1) The atomic ($\sim\mathrm{\AA}$) scale, which defines the electronic structure of the charge-carrying quasi-particles (electrons and holes), intrinsic to the material; (2) the device scale ($\sim\mathrm{nm}$), determined by extrinsic factors such as applied fields, contacts and doping.
For our purposes, we assume that the complex quasi-particle dynamics in a device is well-described by an effective single-particle Schr\"odinger equation of the form,
\begin{equation}
  - \frac{\hbar^2}{2m} \nabla^2 \psi(\vr)
  + \big[ V^\mathrm{c}(\vr) + V^\mathrm{e}(\vr) \big] \psi(\vr)
  = E \psi(\vr)
  \,, \label{e:schroedinger}
\end{equation}
where $V^\mathrm{c}(\vr)$ describes the intrinsic crystal potential, and the extrinsic potential $V^\mathrm{e}(\vr)$ captures the variations of the potential at the device length-scale.
In our case, the crystal potential is given by local atomistic empirical pseudopotentials of each atom $\alpha$,
\begin{equation}
    V^\mathrm{c}(\vr) = \sum_\alpha V^\alpha(|\vr - \vR_\alpha|)\,,
    \label{e:local_epp}
\end{equation}
where $V^\alpha(r)$ represents the radial empirical pseudopotential of atom $\alpha$, centered at location $\vR_\alpha$.
As will be highlighted later on, our method is not limited to this specific form of the crystal potential, and could be extended to non-local, and even \textit{ab-initio} pseudopotentials.
However, in this paper, we will limit our discussion to local empirical pseudopotentials of the form specified in Eq.~(\ref{e:local_epp}).

Various existing computational models discretize Eq.~(\ref{e:schroedinger}) by introducing an appropriate basis-set to capture the smallest atomic scale.
In tight-binding (TB) methods, a limited set of atomic orbitals is used to capture the atomic scale, while on-site potential variations are used to capture the extrinsic potential.~\cite{Mehl1996,Fonseca2013,Huang2012}
In plane-wave based pseudopotential methods, the envelope-function approach has been used to capture the extrinsic potential variations.~\cite{Fischetti2011b,Fang2016}
In both approaches, the total Hamiltonian in Eq.~(\ref{e:schroedinger}), including the extrinsic potential that varies only at the device scale, is solved on the basis set that is used to capture the small atomic scale (atomic orbitals or plane-waves).
This is acceptable for the TB method that scales linearly and features a small basis set of $N_\mathrm{orbitals}$ and $\mathcal{O}(N_\mathrm{bands} N_\mathrm{orbitals})$ complexity, thanks to their nearest-neighbor interactions.~\cite{Fonseca2013,Huang2012}
Plane wave methods, on the other hand, are severely restricted by their large number of plane waves ($N_\vec{G}$) that scales with the volume of the structure rather than the number of electron.
Efficient plane-wave methods, using the Fast Fourier Transform (FFT) algorithm, reduce the complexity of plane-wave algorithms to $\mathcal{O}(N_\mathrm{bands} N_\vec{G} \log N_\vec{G})$, albeit with a rather large pre-factor.~\cite{Kresse1996,VandePut2016}
However, the lack of periodic boundary conditions in the transport direction ($z$), induced by the extrinsic potential, prohibits the use of the FFT algorithm in the transport direction, increasing the complexity to $\mathcal{O}(N_\mathrm{bands} [ N_{G_z}^2 + N_\vec{G} \log N_{\vec{G}_{xy}}])$.
The large basis set, combined with sub-optimal scaling, necessitates a different approach for transport calculations that use plane-wave pseudopotentials.

Instead of treating the intrinsic crystal Hamiltonian and the extrinsic potential with a single method, we propose an alternative approach, where the atomic and device scales are decoupled.
First, we determine the Bloch wave solutions of the intrinsic crystal Hamiltonian, and in a second step, we solve the Hamiltonian of the entire device.
This approach allows us to simulate systems that are currently inaccessible using plane-wave based atomistic pseudopotentials.~\cite{Fang2016,Fang2017}

\begin{figure}[htbp]
    \centering
    \includegraphics{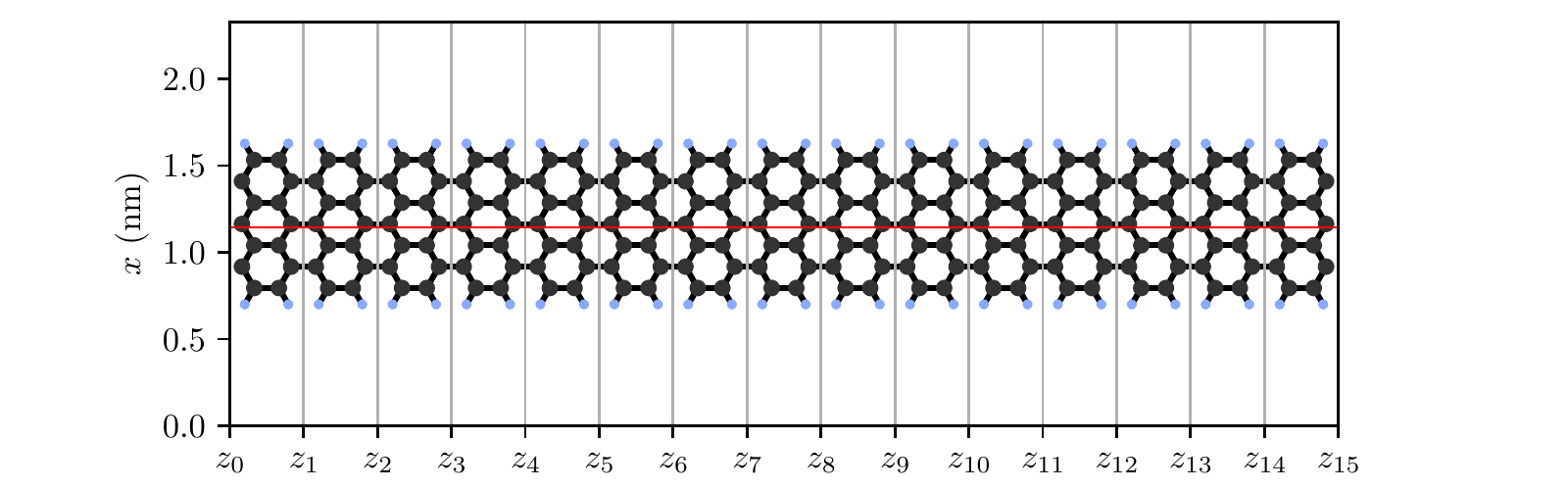}
    \caption{A top-view of an armchair graphene nanoribbon, where carbon (black) and hydrogen (blue) atom positions are indicated with spheres and where black lines represent chemical bonds. Electron transport proceeds in the $z$-direction, where node positions $z_i$ indicate the boundaries between repeated supercells.}%
    \label{f:structure}
\end{figure}

Figure~\ref{f:structure} shows a typical target structure, featuring one-dimensional electron transport, which is assumed to be in the $z$-direction.
The structure consists of a supercell that is periodically repeated $N_\mathrm{block}$ times in the transport direction.
The periodic supercell completely captures the atomic configuration of the one-dimensional crystal.
Extensions to inhomogeneous systems, where the supercell changes throughout the structure are possible, but left for future work.

\subsection{Bloch-wave expansion}

\begin{figure}[htbp]
    \centering
    \includegraphics{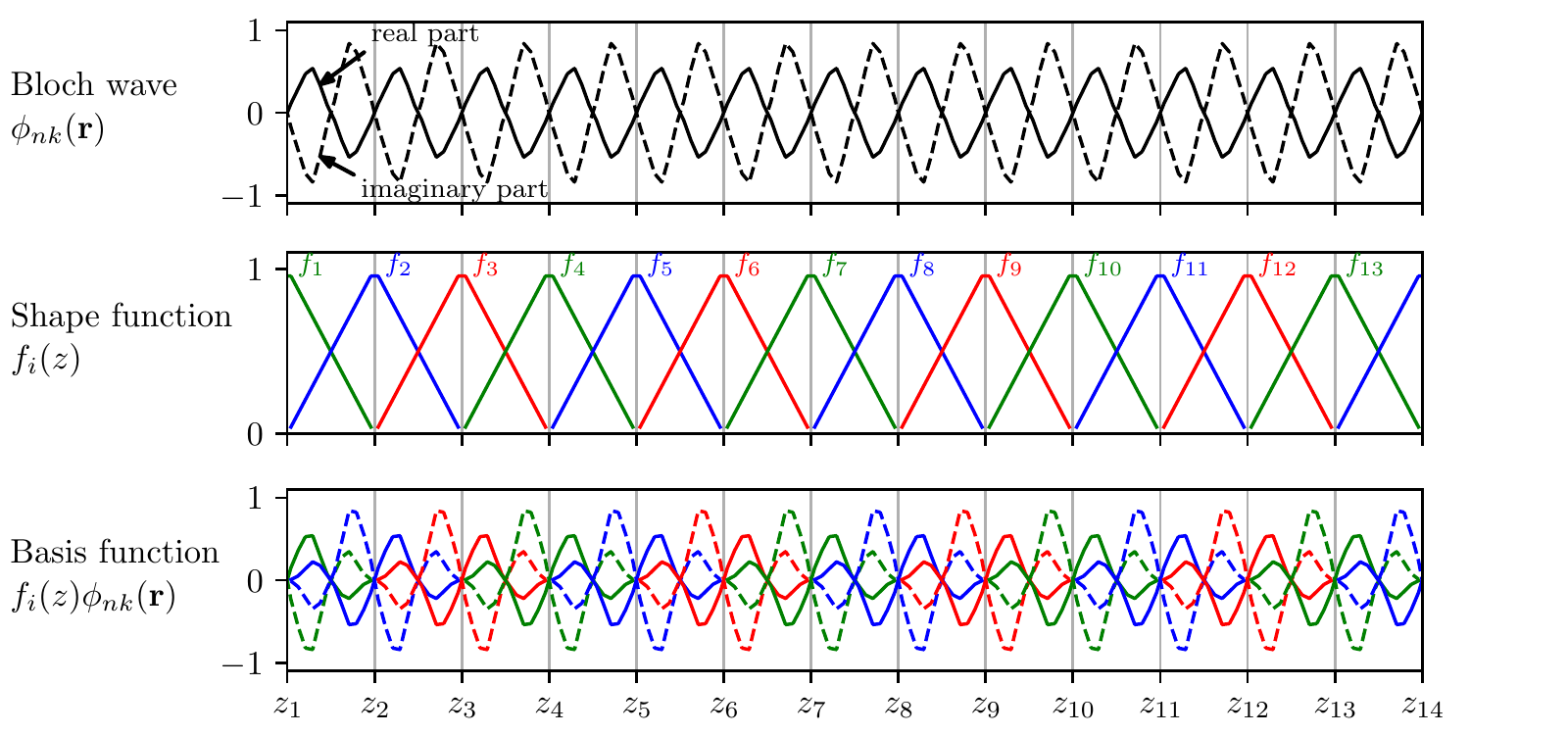}
    \caption{An illustration of the components of the basis set used to expand the wavefunction for the armchair graphene nanoribbon shown in Fig.~\ref{f:structure}. The Bloch wave of the 32nd band ($n=31$) at the $\Gamma$-point ($k=0$) is shown along a cut-line through the middle of the ribbon. The triangular shape functions, forming a partition of unity, are shown for all nodes. The local basis functions that are shown correspond to the Bloch wave in the first panel, \ie, $n=31$ and $k=0$, and are plotted along the same cut-line. The Bloch-wave and basis-function units are arbitrary.}%
    \label{f:wavef_expansion}
\end{figure}

Our method is constructed around an expansion of the wavefunctions on a Bloch-wave basis.
At a high level, our method proceeds as follows: we separate the device in its supercells, we calculate the Bloch waves in each supercell and ``stitch'' them together using finite-elements.
Figure~\ref{f:wavef_expansion} shows the different ingredients for the basis, which we detail in this section.

The first ingredient of our method are the Bloch-waves of the atomic structure, as illustrated in the first panel of Fig.~\ref{f:wavef_expansion}.
For a single repeated supercell, we compute the solution of the intrinsic crystal Hamiltonian with periodic boundaries,
\[
    \left[ -\frac{\hbar^2}{2m} \nabla^2 + V^\mathrm{c}(\vr)\right]
      \left[ u_{nk}(\vr) e^{\ii k z} \right]
    = \epsilon_{nk}\, u_{nk}(\vr) e^{\ii k z}\,.
\]
The solutions are the Bloch functions $u_{nk}(\vr) e^{\ii k z}$, with band index $n$ and wave vector $k$ in the direction of transport.
The Bloch-wave solutions are obtained to high precision using the appropriate plane-wave basis, where computational efficiency is realized using FFTs.~\cite{VandePut2016}

The second ingredient is a one-dimensional finite element (FE) discretization in the transport direction which will ``stitch'' together the supercells and allow for the capture of any extrinsic fields.
The finite element discretization uses the supercells as elements, with nodes $z_i$ located on the interface between the supercells along the transport direction, as shown in Fig.~\ref{f:structure}.
The FE shape functions $f_i(\vr)$, as shown in the second panel of Fig.~\ref{f:wavef_expansion}, are the standard linear FE `hat' shape functions which obey $f_i(\vr_j) = \delta_{ij}$.

The last panel of Figure~\ref{f:wavef_expansion} shows the product of the FE shape functions $f_i(\vr)$ and the node-centered Bloch-waves, defined as:
\begin{equation}
    \phi_{ink}(\vr) = u_{nk}(\vr) e^{\ii k(z - z_i)}\,.
\end{equation}
The products $f_i(\vr) \phi_{ink}(\vr)$ form the Bloch-wave basis-functions on which the wavefunction is expanded,
\begin{equation}
    \psi(\vr)
    = \sum_{ink} c_{ink}\, f_i(\vr) \phi_{ink}(\vr)\,.
    \label{e:wavef_expansion}
\end{equation}
The shape functions $f_i(\vr)$ capture the overall, global variation of the wavefunction, much like the slowly varying envelope functions commonly used.
Note that the shape functions $f_i(\vr)$ also serve to localize the basis functions within the two elements around the node.
The explicit inclusion of the wave vector $k$ in the node-centered Bloch-waves allows for the expansion on more than one (high-symmetry) point of the reciprocal lattice, as will be shown later.

The expansion presented in Eq.~(\ref{e:wavef_expansion}) is a specific application of the Partition-of-Unity Method (PUM).~\cite{Babuska1996,Strouboulis2000,Babuska2004}
In the PUM, a set of overlapping patches $\{\Omega_i\}$ is defined which form an open cover of the complete coordinate space $\Omega$, covering the device.
In our case, a patch $\Omega_i$ is defined as the union of the two supercells touching the node $z_i$.
Adopting the PUM terminology, a shape function $f_i(\vr)$ takes on the role of a patch function that is only supported on the patch $\Omega_i$.
The set of patch (shape) functions $\{f_i(\vr)\}$ satisfies $\forall \vr\in\Omega: \sum_i f_i(\vr) = 1$ and is therefore called a partition-of-unity on the full domain $\Omega$.
The PUM allows for the further enhancement of each patch with a set of functions $\{\phi_{ink}(\vr)\}$ that span an appropriate subspace of the solution space on the patch $\{\phi_{ink}(\vr) | \phi_{ink}(\vr) \subset \mathrm{H}^1(\Omega_i)\}$.
In other words, the linear combination of $\phi_{ink}(\vr)$ should be a good approximation of the solution on the patch.
In our case, the node-centered Bloch-waves $\phi_{ink}(\vr)$ take on the role of enhancement functions, capturing the solution on the atomic scale within the supercell.
The wavefunction function is then well approximated in the solution space of the full domain $\{\psi(\vr) | \psi(\vr) \subset \mathrm{H}^1(\Omega)\}$ by an expansion on the patches, as defined in Eq.~(\ref{e:wavef_expansion}), where the expansion coefficients $c_{ink}$ are to be determined numerically.
Note that the partition of unity formed by $f_i(\vr)$ enforces continuity of the solution independent of the enhancement functions (node-centered Bloch-waves) $\phi_{ink}(\vr)$.
  
\subsection{Matrix equations}

Inserting the expression for the wavefunction in Eq.~(\ref{e:wavef_expansion}), into the Sch\"odinger equation Eq.~(\ref{e:schroedinger}), we determine a linear system of equations for the expansion coefficients $c_{ink}$.
Following the Galerkin method, we convert the Sch\"odinger equation into a weak form, multiplying it by a test function $\bar\psi(\vr)$ and integrating it over the full domain $\Omega$,
\begin{equation}
    - \frac{\hbar^2}{2m} \int_\Omega\rd^3r\, \bar\psi(\vr) \mathrm{H}^\mathrm{(c)} \psi(\vr)
    + \int\rd^3r\, \bar\psi(\vr) [V^\mathrm{e}(\vr) - E] \psi(\vr)
    = 0 \,.
    \label{e:weak_schroedinger}
\end{equation}
This weak form is equivalent to the Schr\"odinger equation when the test functions $\bar\psi(\vr)$ span the full solution space.
The complex conjugate of the wavefunctions form a natural choice for the test functions in Eq.~(\ref{e:weak_schroedinger}).
After expansion, Eq.~(\ref{e:weak_schroedinger}) becomes
\begin{equation}
    \sumall \bar{c}_{i'n'k'} \left[
        \mathrm{H}_{i'n'k',ink}^\mathrm{c}
        + \mathrm{V}_{i'n'k',ink}^\mathrm{e}
        - E\, \mathrm{M}_{i'n'k',ink}
    \right] 
    c_{ink}
    = 0 \,,
    \label{e:weak_w_matel}
\end{equation}
where we have introduced the matrix elements
\begin{align}
    \mathrm{M}_{i'n'k',ink} 
    &= \intall f^*_{i'}(\vr) \phi^*_{i'n'k'}(\vr) 
               f_i(\vr) \phi_{ink}(\vr) \,,
    & \text{(overlap / ``mass'')} \\
    \mathrm{H}_{i'n'k',ink}^\mathrm{c}
    &= \intall f^*_{i'}(\vr) \phi^*_{i'n'k'}(\vr)
               \mathrm{H}_\mathrm{c}(\vr) 
               f_i(\vr) \phi_{ink}(\vr) \,,
    & \text{(crystal Hamiltonian)} \label{e:crystal_ham_matel} \\
    \mathrm{V}_{i'n'k',ink}^\mathrm{e}
    &= \intall f^*_{i'}(\vr) \phi^*_{i'n'k'}(\vr)
               V_\mathrm{e}(\vr) 
               f_i(\vr) \phi_{ink}(\vr) \,.
    & \text{(extrinsic potential)}
\end{align}
Note that using the complex conjugates of the Bloch basis as the test functions preserves the Hermiticity of the discretized Hamiltonian and overlap matrices.

The direct evaluation of the crystal Hamiltonian matrix elements in Eq.~(\ref{e:crystal_ham_matel}) requires the use of the crystal potential.
While this is fairly easy for the case of the local empirical pseudopotential approximation, the evaluation of the crystal Hamiltonian in, \eg, \textit{ab-initio} methods, can be more cumbersome or computationally expensive.
To make our model independent of the intricacies to evaluate the crystal Hamiltonian, we avoid the direct use of the crystal potential itself by substituting the eigenvalues of the crystal Hamiltonian in Eq.~(\ref{e:crystal_ham_matel}) (the full details are given in~\ref{a:weak_bloch}),
\begin{equation}
    \mathrm{H}^\mathrm{c}_{i'n'k',ink} 
    = \frac{\epsilon_{ink} + \epsilon_{i'n'k'}}{2}\, \mathrm{M}_{i'n'k',ink}
    + \mathrm{T}_{i'n'k',ink}
    + \mathrm{P}_{i'n'k',ink}\,,
\end{equation}
where $\epsilon_{ink}$ is the eigenvalue of the corresponding Bloch wave $\phi_{ink}(\vr)$ and two new matrix elements have been defined as:
\begin{align}
    \mathrm{T}_{i'n'k',ink}
    &= \frac{\hbar^2}{4m}
       \intall
         \vnabla \big[ f^*_{i'}(\vr) f_i(\vr) \big]
         \cdot
         \vnabla \big[ \phi^*_{i'n'k'}(\vr) \phi_{ink}(\vr) \big]\\
    &+ \frac{\hbar^2}{2m}
       \intall 
         \big[ \vnabla f^*_{i'}(\vr) \big]
         \phi^*_{i'n'k'}(\vr) \cdot
         \big[ \vnabla f_i(\vr) \big] 
         \phi_{ink}(\vr)
      \,,
       & \text{(kinetic energy)} \\
    \mathrm{P}_{i'n'k',ink} 
    &= -\frac{\hbar^2}{m}
       \intall f^*_{i'}(\vr) \phi^*_{i'n'k'}(\vr) 
               \big[ \vnabla f_i(\vr) \big] \cdot
               \big[ \vnabla \phi_{ink}(\vr) \big]
       + \mathrm{h.c.}\,,
       & \text{(momentum coupling)} 
\end{align}
where {h.c.} has been used to indicate the Hermitian conjugate of the previous term, swapping indices $ink$ and $i'n'k'$.
 
Since Eq.~(\ref{e:weak_w_matel}) has to hold for all test functions, \ie, all coefficients $c^*_{ink}$, we write,
\begin{equation}
    \sum_{ink} \left[
      T_{i'n'k',ink}
      + V_{i'n'k',ink}
      + \frac{(\epsilon_{ink} + \epsilon_{i'n'k'})}{2} M_{i'n'k',ink}
      + P_{i'n'k',ink}
    \right] c_{ink}
    = E \sum_{ink} M_{i'n'k',ink}\, c_{ink} \,.
    \label{e:expansion_matrix_eq}
\end{equation}
This generalized eigenvalue problem can be written in matrix form as $\mathrm{H} \vec{c} = E \mathrm{M} \vec{c}$.
Note that, apart from the extrinsic potential, all the matrix elements depend only on the properties of the material, not on those of the device, and are independent of changes of the extrinsic potential.
Thanks to the shape functions, only elements for which $i$ and $i'$ are equal or refer to nearest-neighbor nodes are non-zero.
The matrices $\mathrm{H}$ and $\mathrm{M}$ have a block tridiagonal form.
For example, the Hamiltonian matrix is written as:
\begin{equation}
    \mathrm{H} 
    = \left[\!\!\begin{array}{lllllll} 
        \ddots & & & & & & \iddots \\
        & \mathrm{H}_{i-1,i-2} & \mathrm{H}_{i-1,i-1} & \mathrm{H}_{i-1,i} & 0 & 0 & \\
        & 0 & \mathrm{H}_{i,i-1} & \mathrm{H}_{i,i} & \mathrm{H}_{i,i+1} & 0 &  \\
        & 0 & 0 & \mathrm{H}_{i+1,i} & \mathrm{H}_{i+1,i+1} & \mathrm{H}_{i+1,i+2} & \\
        \iddots & & & & & & \ddots
    \end{array}\!\!\right]%
    \label{e:infinite_system}
\end{equation}
where each block $\mathrm{H}_{ii'}$ (and $\mathrm{M}_{ii'}$ for the overlap matrix) is a square matrix with size equal to the number of basis functions used in a supercell $N_\mathrm{basis}$. Correspondingly, the solution vector $\vec{c}$ combines the column vectors $\vec{c}_i$ that contain the expansion coefficients for slice $i$.

\section{Open system}%
\label{s:transport}

Having obtained a suitable discretization of the atomic structure, we now turn to the calculation of the electronic transport properties in devices.
We consider an open system with injecting and absorbing contacts on either side of the device, here referred to as source (s) and drain (d).
Both contacts are considered infinite reservoirs which inject electrons in thermodynamic equilibrium and absorb all incident waves.
We employ the quantum transmitting boundary condition method (QTBM)~\cite{Lent1990} to model the contacts and calculate the extended states that are injected from each contact.

\subsection{Contact self-energies}

The calculation of contact self-energies using iterative and direct approaches (as used here) is already well established in literature.~\cite{Sorensen2009,Tsukamoto2017,Huang2012,Sorensen2008}
Nonetheless, we will detail the procedure here. 
Our reasons for this are twofold; (1) our basis, being non-orthogonal, introduces additional complexity that, to our knowledge, has not been previously described for the direct approach, and (2) our numerical approach avoids some numerical errors in calculating the self-energies directly.
We note that this procedure can be applied to calculate the self-energies for other non-orthogonal bases, for example in non-orthogonal Gaussian-type tight-binding~\cite{Mehl1996} and projector-augmented wave methods~\cite{Blochl1994}.

We calculate the self-energies $\Sigma_\mathrm{s/d}$, associated with the truncation of the block matrices in Eq.~(\ref{e:infinite_system}) at the open contacts using a direct, non-iterative, method.
For this purpose, we calculate the so-called complex band structure at the source or drain node $i \in\{\mathrm{s,d}\}$, for a given energy $E$, as the solution of the non-linear eigenvalue problem
\begin{equation}
    \big[ \mathrm{H}_i(\lambda) - E \mathrm{M}_i(\lambda) \big] \vec{c}_i = 0 \,,
    \label{e:complex_bs}
\end{equation}
where the eigenvalues $\lambda = \ee^{\ii k\, \Delta z}$ are the phase difference between the edge node $i$ and its nearest neighbor inside the contact $i+1$ for the drain ($i-1$ for the source), with $\Delta z = z_{i} - z_{i+1}$.
The polynomial matrices are given by
\begin{equation}
    \mathrm{H}_i(\lambda) 
    = \lambda^{-1} \mathrm{H}_{i,i-1} + \mathrm{H}_{i,i} + \lambda \mathrm{H}_{i,i+1}
    \quad\mathrm{and}\quad
    \mathrm{M}_i(\lambda)
    = \lambda^{-1} \mathrm{M}_{i,i-1} + \mathrm{M}_{i,i} + \lambda \mathrm{M}_{i,i+1}\,.
\end{equation}
Equation~(\ref{e:complex_bs}) represents a second-order, generalized eigenvalue equation.
This can be solved readily by linearizing the second-order eigenvalue problem to a first order problem of double the rank.
To avoid excessive numerical round-off errors in the calculation of the eigenvalues $\lambda$, we linearize Eq.~(\ref{e:complex_bs}) using the symmetric scheme from Ref.~\cite{Higham2005},
\begin{equation}
    \begin{bmatrix}
        \mathrm{H}_{i,i} - E \mathrm{M}_{i,i} & \mathrm{H}_{i,i-1} - E \mathrm{M}_{i,i-1} \\
        \mathrm{H}_{i,i+1} - E \mathrm{M}_{i,i+1} & 0
    \end{bmatrix}
    \begin{bmatrix}
        \vec{d}_i \\ \vec{c}_i
    \end{bmatrix}
    = \lambda
    \begin{bmatrix}
        - \left( \mathrm{H}_{i,i+1} - E \mathrm{M}_{i,i+1} \right) & 0 \\
        0 & \mathrm{H}_{i,i+1} - E \mathrm{M}_{i,i+1}
    \end{bmatrix}
    \begin{bmatrix}
        \vec{d}_i \\ \vec{c}_i
    \end{bmatrix}\,,
    \label{e:linearized_eigenvalue_symmetric}
\end{equation}
where $\vec{d}_i = \lambda \vec{c}_i$, and the left-hand-side is a Hermitian matrix, since $\mathrm{H}_{i,i+1} = \mathrm{H}_{i,i-1}^\dagger$ and $\mathrm{M}_{i,i+1} = \mathrm{M}_{i,i-1}^\dagger$.

Equation~(\ref{e:linearized_eigenvalue_symmetric}) is solved to machine precision using a direct linear eigenvalue solver and admits $2 N_\mathrm{basis}$ solution pairs $(\lambda_\nu, \vec{c}_\nu)$.
Based on the phase factors $\lambda_\nu$, we sort them into two sets of size $N_\mathrm{basis}$ each, the in-flowing and out-flowing solutions.
To determine flow-direction, we calculate the group velocity $v_\nu$ of each eigenvector $\vec{c}_\nu$ using a generalization of the Hellmann-Feynman theorem~\cite{Feynman1939},
\begin{equation}
    v_\nu
    = \frac{1}{\hbar} \pd{E_\nu}{k}
    = \frac{1}{\hbar} 
      \frac{\matrixel{\vec{c_\nu}}{\pd{}{k}\mathrm{H}_i(\lambda_\nu)}{\vec{c}_\nu} 
              - E_\nu \matrixel{\vec{c_\nu}}{\pd{}{k}\mathrm{M}_i(\lambda_\nu)}{\vec{c}_\nu}}
           {\matrixel{\vec{c_\nu}}{M_i(\lambda_\nu)}{\vec{c}_\nu}} \,.
   \label{e:hellmann-feynman}
\end{equation}
The set of solutions with an out-flow (in-flow) condition is split into purely traveling waves with $|\lambda_\nu|=1$ and $v_\nu > 0$ ($v_\nu < 0$), and evanescent modes where $|\lambda| < 1$ ($|\lambda| > 1$).
In practical implementations, a tolerance should be used to determine the traveling waves, \ie, $|\lambda_\nu|=1 \pm \varepsilon$.
Thanks to the increased accuracy of the symmetric linearization of Eq.~(\ref{e:complex_bs}), we obtained a drastic improvement in the accuracy of $|\lambda_\nu|$ and all traveling modes satisfy $|\lambda|=1$ to machine precision in all our tests.
In the envelope-function approximation, a necessary additional step is the removal of spurious solutions.~\cite{Fang2016}
However, our method does not admit spurious traveling solutions within (or below) the energy range spanned by the Bloch waves in the basis set, negating the need for additional filtering.

Before proceeding, care must be taken to correctly normalize the traveling wavefunctions in each contact.
In the infinitely long contacts, the wavefunctions for different values of the crystal momentum $k_z$ are orthonormal, $\int_{\Omega_\mathrm{s/d}}\rd^3r\,\psi_{k_z}^*(\vr) \psi_{k_z'}(\vr) = \delta(k_z-k_z')$, where the domain $\Omega_\mathrm{s/d}$ spans the entire infinite contact.
When the integration domain is reduced to a single supercell $\Omega_\mathrm{sc}$, the normalization condition for the wavefunction becomes
\begin{equation}
	\int_{\Omega_\mathrm{sc}}\rd^3r\, \psi_{k_z}^*(\vr) \psi_{k_z}(\vr)
	= \frac{L_z}{2\pi}\,,
\end{equation}
where $L_z$ is the length of the supercell along the transport direction.
In terms of our wavefunction expansion, the condition is straightforward:
\begin{equation}
    \matrixel{\vec{c}_\nu}{\mathrm{M}_i(\lambda_\nu)}{\vec{c}_\nu} = \frac{L_z}{2\pi}\,.
	\label{e:normalization}
\end{equation}
This normalization condition is applied immediately upon identification of the running modes we obtain after solving the complex band structure in Eq.~(\ref{e:complex_bs}).

For each contact node $i\in\{\mathrm{s,d}\}$, we define a Bloch matrix, $\mathrm{B}_i = [\vec{c}_1, \ldots,\vec{c}_\nu, \ldots, \vec{c}_{N}]$, whose columns are the out-flow eigenvectors $\vec{c}_\nu$ of the respective contact.
The contact self-energy of the contact-node $i\in\{\mathrm{s,d}\}$ is built by projecting the wavefunction in the device on the out-flowing waves,
\begin{equation}
    \Sigma_i
    = \big[ \mathrm{H}_i^{\prime} - E \mathrm{M}_i^{\prime} \big]
      \mathrm{B}_{i} \Lambda \mathrm{B}_{i}^{-1} \,,
\end{equation}
where $\Lambda_{i,\mathrm{out}}$ is a diagonal matrix with elements given by the out-flow $\lambda_i$, while the $N_\mathrm{basis}\times N_\mathrm{basis}$ matrices $\mathrm{H}_i^{\prime}$ and $\mathrm{M}_i^{\prime}$ correspond to the truncated matrices just outside the simulation domain, \eg, $\mathrm{H}_i^\prime = \mathrm{H}_{i,i-1}$ for the source contact.
The effect of the projection can be understood as follows: $\mathrm{B}^{-1}$ converts the wavefunction into the coefficients of each mode, $\Lambda$ propagates the coefficients to the next node by multiplying each mode with $e^{\ii k_z\Delta z}$ and $\mathrm{B}$ converts the coefficients back into its wavefunction form.
Finally, we define $\Sigma$, a matrix of the size of the system ($N_\mathrm{basis}N_\mathrm{block}\times N_\mathrm{basis}N_\mathrm{block}$) that contains the two contact self-energy matrices $\Sigma_\mathrm{s}$ and $\Sigma_\mathrm{d}$ at their respective positions on the diagonal, and is zero otherwise.

\subsection{Extended states}

Using the contact self-energies, we calculate the extended states of the open system by solving directly for the coefficients $c_{ink}$ of the wavefunction,
\begin{equation}
    \big[ E \mathrm{M} - \mathrm{H} - \Sigma \big] \vec{c} = \mathrm{B}\,,
    \label{e:qtbm_system}
\end{equation}
where the right-hand-side matrix $\mathrm{B}$ has $N_\mathrm{mode}$ columns that each represent the injection of a single eigenmode from one of the contacts.
For each in-flowing mode $\gamma$ in each contact node $i\in\{\mathrm{s,d}\}$, with coefficients $\vec{c}_{i,\gamma}$ and phase $\lambda_{i,\gamma}$, we obtain
\begin{equation}
    \mathrm{B}_{i,\gamma}
    = \big[ (\mathrm{H}_i^{\prime} - E \mathrm{M}_i^{\prime}) \lambda_{i,\gamma}
            - \Sigma_i \big] \vec{c}_{i,\gamma} \,,
\end{equation}
with $\mathrm{B}$ zero everywhere else.
Having calculated the coefficients for all injected modes $\gamma$ from all contacts by solving Eq.~(\ref{e:qtbm_system}) at a certain energy, the expansion in Eq.~(\ref{e:wavef_expansion}) is used to express the wavefunctions in the real-space basis:
\begin{equation}
    \psi_{\gamma}(\vr) = \sum_{ink} c_{\gamma,ink} f_i(\vr) \phi_{ink}(\vr)\,.
\end{equation}
The label $\gamma$ is used to identify both the originating contact (s/d) and individual injected mode index.

Rather than following the procedure described above, we could also use the popular nonequilibrium Green's function (NEGF) approach and solve for the Green's function in our Bloch wave basis $\mathrm{G} = {\left[ E \mathrm{M} - \mathrm{H} - \Sigma \right]}^{-1}$.
NEGF can be implemented efficiently by using an appropriate recursive technique, calculating only the diagonals and off-diagonals of the Green's function~\cite{Li2012,Kuzmin2013,Kazymyrenko2008}.
Such a recursive Green's function approach would, in our case, reduce the computational complexity from the inversion of the entire Hamiltonian, $\mathcal{O}(N_\mathrm{blocks}^2\times N_\mathrm{basis}^2)$, to the inversion of the individual blocks of size $N_\mathrm{basis}$, \ie, $\mathcal{O}(N_\mathrm{blocks}\times N_\mathrm{basis}^2)$.
However, in general, the number of traveling modes $N_\mathrm{mode}$ using wavefunctions is much smaller than the number of basis vectors $N_\mathrm{basis}$ at a single node.
Therefore the QTBM based on wave functions, with a complexity of $\mathcal{O}(N_\mathrm{blocks}\times N_\mathrm{basis} \times N_\mathrm{modes})$, is more efficient than solving for the Green's function, as already noted by Bruck \textit{et al.}~\cite{Bruck2017}.
Both approaches are identical when considering ballistic transport.~\cite{Milnikov2012}

\subsection{Density}%

The full electron density of the open system is formally given by
\begin{equation}
    n(\vr) 
    = \int\rd E  \sum_\nu g_\nu(E) |\psi_{E\nu}(\vr)|^2 f_\mathrm{FD}(E - \mu_\nu)\,,
    \label{e:density_3d}
\end{equation}
where $g_\nu(E)$ represents the density of states (including spin degeneracy) of the injecting contact of mode $\nu$, calculated from the velocity determined from the generalized Hellmann-Feynman theorem (Eq.~(\ref{e:hellmann-feynman})), and $f_\mathrm{FD}(E - \mu_\nu)$ is the Fermi-Dirac distribution, where $\mu_\nu$ is the electrochemical potential in the contact of mode $\nu$.
In the evaluation of the integral over energy $E$, singularities of the type $1/\sqrt{E-E_\mathrm{singularity}}$ are encountered in the density of states at local band-extrema, \ie, where $\rd E/ \rd k = 0$. 
Since the location of these singularities are \textit{a-priori} unknown and the evaluation of the wavefunctions $\psi_{E\nu}(\vr)$ is computationally expensive, we have adopted an adaptive Simpson technique for the numerical evaluation of the integral to a specified numerical tolerance.
In our tests, the Simpson method provides an accurate error estimate, which gives a reliable accuracy for our results.

In a naive implementation of Eq.~(\ref{e:density_3d}), the wavefunctions $\psi_{E\nu}(\vr)$ are evaluated directly using the expansion defined in Eq.~(\ref{e:wavef_expansion}).
This step is computationally expensive, as the Bloch-wave grid, with $N_\vr$ points, is generally very fine. 
However, during the adaptive Simpson integration, we can compute an estimated average density on the $N_\mathrm{nodes}$ nodes, instead of on all $N_\mathrm{nodes}\times N_\vr$ points in space. 
We call this the node density,
\begin{equation}
    \langle n \rangle^\mathrm{node}[z_i] = \int \rd E \sum_\nu \langle n \rangle^\mathrm{node}_{E,\nu}[z_i] f_\mathrm{FD}(E - \mu_\nu)\,,
\end{equation}
where the local density of states of the nodes is simply given by
\begin{equation}
    \langle n \rangle^\mathrm{node}_{E,\nu}[z_i] = g_\nu(E) \sum_{nk} |c_{ink}|^2\,.
\end{equation}
This evaluation of the node density comes at virtually no cost.
Note that to interpret $\langle n \rangle_\mathrm{node}[z_i]$ as an estimate of average of the real density, the Bloch-waves need to be normalized in a specific way,
\begin{equation}
    \int_{\Omega_\mathrm{sc}}\rd^3r\, |u_{nk}|^2 = V_\mathrm{sc}\,,
\end{equation}
where $\Omega_\mathrm{sc}$ covers the supercell and $V_\mathrm{sc}$ is its volume.
With this normalization, the coefficients $c_{ink}$ have units $[\sqrt{m^{-2}}]$ and the normalized Bloch-waves are dimensionless weights that average to unity in a single cell.
Since all coefficients $c_{ink}$ are normalized with respect to the mass matrix $\mathrm{M}$ upon injection (see Eq.~(\ref{e:normalization})), the wavefunctions remain properly normalized.

By storing all the integration energies $E$, weights $w_{E\nu}$ and coefficients $c_{E\nu,ink}$ when computing the node density, we can efficiently reconstruct the complete real-space density in one step, avoiding the costly naive evaluation of Eq.~(\ref{e:density_3d}).
To achieve this, we compute the matrix elements of the density matrix, expressed in the Bloch-basis:
\begin{equation}
    n_{i'n'k',ink} 
    = \sum_{E, \nu} w_{E\nu} g_{E\nu} c^*_{E\nu,i'n'k'} c_{E\nu,ink} f_\mathrm{FD}(E - \mu_\nu)\,.
    \label{e:density_matrices}
\end{equation}
Thanks to the locality of the shape functions in the Bloch-basis, only the matrix-elements for $i=i'$ and nearest neighbors $i,i'$ need to be computed for the evaluation of the density
\begin{equation}
    n(\vr) = \sum_{i'n'k',ink} n_{i'n'k',ink}(\vr) \psi^*_{i'n'k'}(\vr) \psi_{ink}(\vr)\,.
    \label{e:density_3d_fast}
\end{equation}
Careful analysis of the operations involved in each procedure show that the matrix-based approach is faster if the number of Bloch-waves used in the basis is lower or equal to the number of injected states, \ie, $N_\mathrm{basis} < N_\mathrm{waves}$.

\begin{figure}[!htbp]
    \centering
    \includegraphics{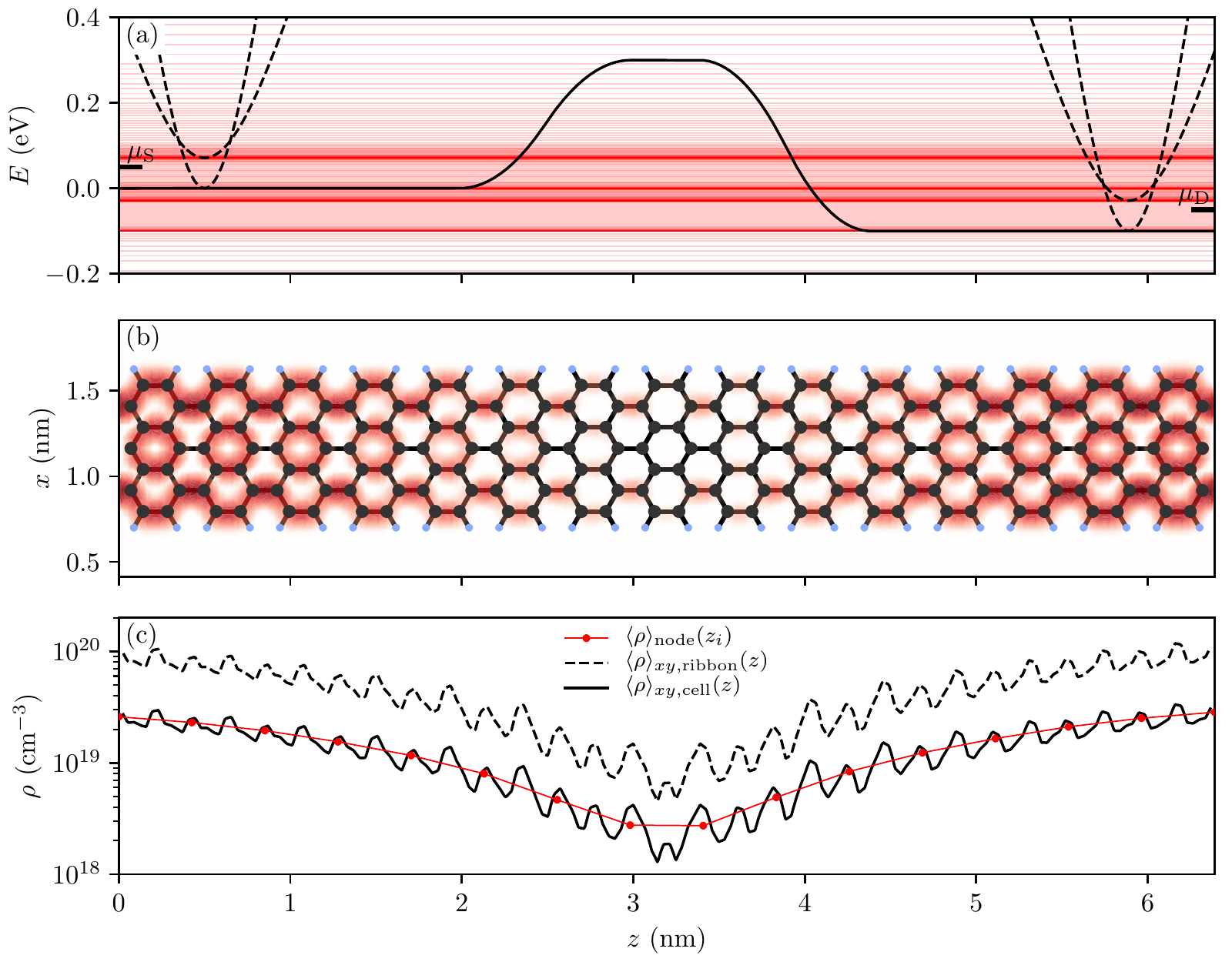}
    \caption{An illustration of the calculation of the density.
        (a) Band-diagram showing the variation of the bottom of the conduction band {w.r.t.} the $z$ coordinate and the chemical potentials of the source and drain contacts $\mu_\mathrm{s/d}$ ($50\,\mathrm{meV}$ above their respective conduction band). Each horizontal (red) line is an energy of injection (289 total), determined by the adaptive Simpson integrator for a tolerance of ${10}^{14}\ \mathrm{cm}^{-3}$. For reference, the band structure in the source and drain regions are shown as dashed lines.
        (b) The resulting free electron density in the ribbon, averaged over the $y$-direction, shown as pseudo-color. 
        (c) The same electron density, averaged over the $x-y$ plane of the full cell ($\langle\rho\rangle_{xy,\mathrm{cell}}$), averaged over an $x-y$ plane inside the ribbon only ($\langle\rho\rangle_{xy,\mathrm{ribbon}}$), and the `node averaged' density ($\langle\rho\rangle_\mathrm{node}$) as explained in the text.}%
    \label{f:density}
\end{figure}
Figure~\ref{f:density} shows an example of the Simpson integration with a fixed potential as shown in Fig.~\ref{f:density} (a).
The selected integration energies are indicated in Fig.~\ref{f:density} (a), showing adaptive refinements near the band extrema of the contacts, as expected.
The converged $y$-averaged density in Fig.~\ref{f:density} (b), clearly shows the sub-atomic resolution of the reconstructed density.
We also show that the node density in Fig.~\ref{f:density} (c) matches very well the $xy$-averaged density inside the ribbon.

Since we use the node density in the Simpson integration, the error estimate that is used for the refinement is based on the node density.
In theory, we can not guarantee that a specified tolerance for the node density, using the error estimate on the node density, is an exact measure of the error of the complete density at every point.
We verify that this is not an issue in practice by determining the accuracy of the calculation of the density in Fig.~\ref{f:density}.
We first request an absolute tolerance of the integrated density of $10^{14}\ \mathrm{cm^{-3}}$ and then perform a more accurate calculation with a tolerance of $10^{11}\ \mathrm{cm^{-3}}$ (three orders of magnitude smaller).
A comparison of the two results shows a root-mean-squared error of $1.4\times10^{13}\ \mathrm{cm^{-3}}$ for the node density, close to the requested value, and a difference of $33\times10^{13}\ \mathrm{cm^{-3}}$ for the complete real-space density.
As expected, the error on the density is underestimated by the error on the node density due to the atomic variations.
In practice, we account for this discrepancy by selecting a node-density tolerance at least two orders of magnitude smaller than the required charge density.

\subsection{Transmission and Ballistic Current}

The transmission probability of each state is calculated by taking the ratio of the injected and transmitted current density.
For a mode $\gamma$ injected from the source contact (s), the transmission probability to the drain contact (d) are calculated from the Bloch matrices and group velocity as
\begin{equation}
    T_\mathrm{sd}
    = \frac{J_\mathrm{d,out}}{J_\mathrm{s,inj}}
    = \frac{\sum_\mu {\left|
            {\left[\mathrm{B}^{-1}_\mathrm{d}\right]}_{\mu\gamma} 
                \vec{c}_{\mathrm{d},\gamma}
        \right|}^2 v_{\mathrm{d},\mu} }
        {v_{\mathrm{s},\gamma}}\,,
	\label{e:transmission}
\end{equation}
assuming the injected coefficients are properly normalized, \ie, $|\vec{c}^\mathrm{inj}_{\mathrm{s},\gamma}| = 1$.
As a sanity check, we explicitly calculate the reflection coefficient $T_\mathrm{ss}$ with Eq.~(\ref{e:transmission}) after first removing the injected part $\vec{c}^\mathrm{inj}_{\mathrm{s},\gamma}$ from the coefficients $\vec{c}_{\mathrm{s},\gamma}$, $T_\mathrm{sd} + T\mathrm{sd} = 1$.

The ballistic current from source (s) to drain (d) is calculated from the transmission coefficients as
\begin{equation}
    I_\mathrm{sd} 
    = \int\rd E \sum_{\nu} g_\nu(E)\, T_\mathrm{sd,E\nu} \,\sgn(v_{E\nu}) f_\mathrm{FD}(E - \mu_\nu) \,,
    \label{e:current}
\end{equation}
where $\sgn(v_{E\nu})$ gives the sign of the velocity of the injected state, \ie, $+1$ ($-1$) for states $\nu$ originating from the source (drain). 
The integral in Eq.~(\ref{e:current}) is evaluated using the same adaptive Simpson method used to calculate the density.
A separate integration of the current, rather than using the states obtained in the density simulation, is advised since the energies that are associated with a high current do not necessarily align with those that are responsible for the density.
Furthermore, since the current integration is free of singularities, the integration converges quickly, with fewer evaluations than required in the density simulation for the same relative accuracy.

\section{Self-consistency}%
\label{s:self-consistent}

In realistic electronic devices, external potentials are applied by gates and fixed charges are associated with ionized doping.
To account for all these effects, as well as the mean-field interaction of the electron charge, we adopt the Hartree approximation.
The extrinsic potential is found self-consistently with the electron density by solving the non-linear Poisson equation,
\begin{equation}
    \vnabla \cdot \left[\epsilon(\vr) \vnabla V(\vr) \right] 
    = \rho[\vr; V] + \rho_\mathrm{doping}(\vr) \,,
    = \rho_\mathrm{net}[\vr; V] \,,
    \label{e:nonlin_poisson}
\end{equation}
where $\rho[\vr; V] = -\ee n[\vr; V]$ is the free-electron charge for a given potential $V(\vr)$, $\rho_\mathrm{doping}(\vr)$ represents the fixed charge density originating from the ionized dopants, and $\rho_\mathrm{net}[\vr; V]$ is the net charge density.

To allow for the general application of boundary conditions and shapes for the gates and doping profiles the density and potential are discretized on a linear tetrahedral finite-elements mesh, forming the Poisson domain $\Omega_\mathrm{Poisson}$.
At the edges of the Poisson domain, \ie, for $\vr\in\partial \Omega_\mathrm{Poisson}$, we impose Neumann boundary conditions $\vnabla V(\vr) \cdot \vec{\hat{n}}(\vr) = 0$, where $\vec{\hat{n}}$ is the normal to the edge.
The electrostatic control of the device by gates is included by applying Dirichlet boundary conditions to their domains.
For a single gate at a fixed potential $V_\mathrm{g}$ with domain $\Omega_\mathrm{g}$ the Dirichlet condition is $V(\vr \in \Omega_\mathrm{g}) = V_\mathrm{g}$.
A high quality tetrahedral mesh, covering the Poisson domain, and conforming to the gates is automatically generated.

However, since the density, constructed using Eq.~(\ref{e:density_3d}), is given on a uniform grid corresponding to the Fourier transform of the plane-wave components of the Bloch waves, the density needs to be interpolated to the tetrahedral finite-elements mesh.
To avoid unnecessary approximation and the introduction of errors, the finite-element mesh explicitly includes all points of the uniform Bloch-wave mesh where the Bloch waves that comprise the basis set have non-negligible values.
Specifically, a point $\vr_l$ from the uniform grid is included in the tetrahedral mesh if 
\begin{equation}
    |u_{nk}(\vr_l)|^2 > {10}^{-3} \times \max_{\{n,k,\vr\}} |u_{nk}(\vr)|^2 \,,
\end{equation}
for any band $n$ and wave-vector $k$ in the basis-set. 
To cover the entire Poisson domain, additional mesh points are generated automatically using the TetGen library.~\cite{Si2015}
This procedure yields a coarser global mesh with a gradual transition to the fine mesh points determined by the Bloch waves.
This way of constructing the mesh is equivalent to an adaptively refined mesh in regions of high (expected) density.
Since the potential is calculated on the tetrahedral mesh, the calculation of the matrix elements of the extrinsic potential $\mathrm{V}_{i'n'k',ink}^\mathrm{e}$ requires an interpolation of the tetrahedral mesh to the uniform Bloch-wave mesh.
By sharing points between the tetrahedral mesh and the uniform mesh, we introduce a significant amount of additional bookkeeping.
However, doing so we combine the sub-atomic resolution of Bloch-waves with the ability of the mesh to comply to general boundary conditions.

In addition to the flexibility in applying boundary conditions, a tetrahedral mesh provides a significant decrease in computational burden compared to the uniform Bloch-wave mesh that could otherwise be used, since the number of points in the tetrahedral mesh is significantly lower than those of the Bloch waves, $N_\mathrm{tetra} \ll N_\vr \times N_\mathrm{blocks}$.
The penalty we incur consists in the burden of interpolation whenever we transition between the meshes.
For this purpose, we use a linear interpolation that matches the linear shape functions used in the finite-element representation of the linear Poisson equation.
However, the interpolation burden is limited because the values on the points that are shared between the two meshes, which account for the majority of points in all our test structures, do not require interpolation.

\begin{figure}[htbp]
    \centering
    \includegraphics{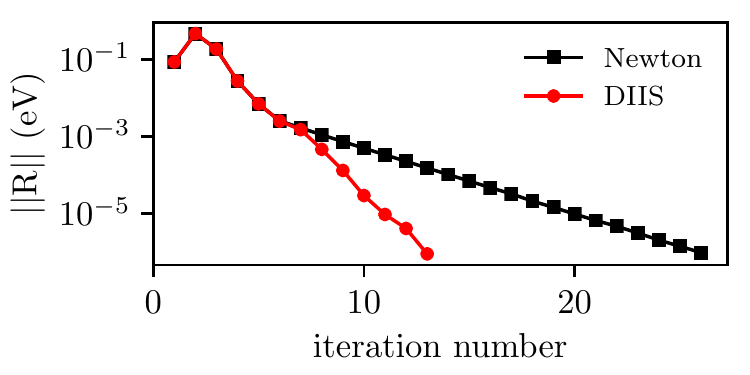}
    \caption{%
        Convergence rate of self-consistent procedure for the device shown in Fig.~\ref{f:device} in the off-state ($V_\mathrm{g}=-0.2\ \mathrm{V}$ and $V_\mathrm{ds}=0.2\ \mathrm{V}$) from a uniform starting potential.
        The $l^2$-norm of the residual is shown for the semi-classical Newton iteration, as well as the accelerated DIIS method.
        The convergence criterion is set to $10^{-6}\ \mathrm{eV}$.
    }%
    \label{f:sc_convergence}
\end{figure}

Using linear shape functions on the tetrahedra, we arrive at the FE representation of the non-linear Poisson equation
\begin{equation}
    \mathrm{D} \vec{V} 
	= \mathrm{M} \left\{ 
		\boldsymbol{\rho}[\vec{V}] + \boldsymbol{\rho}_\mathrm{doping} 
	\right\} 
	= \mathrm{M} \boldsymbol{\rho}_\mathrm{net}[\vec{V}] \,, 
    \label{e:nonlin_poisson_fem}
\end{equation}
where $\mathrm{M}$ is the mass-matrix and $\mathrm{D}$ represents the $\vnabla \cdot \left[\epsilon(\vr) \vnabla \right]$ operator.
The non-linear Poisson equation is solved using a Newton-Rhapson method, which, for iteration $p+1$ reads:
\begin{equation}
	\vec{V}^{p+1} 
	= \vec{V}^{p} 
	- {\left[\mathrm{J}^p\right]}^{-1} \Big[
        \mathrm{D} \vec{V}^p - \mathrm{M} \boldsymbol{\rho}_\mathrm{net}\left[\vec{V}^p\right] 
	\Big]\,,
	\label{e:newton_poisson}
\end{equation}
where the Jacobian is given by $\mathrm{J}^p = \mathrm{D} - \mathrm{M} \mathrm{J}^p_\rho$.
We approximate the Jacobian for the free charge density $\mathrm{J}^p_\rho$ with a semi-classical diagonal matrix, calculated by varying the local chemical potential.
In practice, we calculate it by evaluating the free-density in Eq.~(\ref{e:density_3d}) with the derivative of the Fermi-Dirac distribution instead of the Fermi-Dirac distribution itself, \ie,
\begin{equation}
    \tilde{J}_\rho^{p}(\vr, \vr') 
	= \delta(\vr, \vr') \frac{\delta \rho[\vr, V^{p}(\vr)]}{\delta \mu(\vr)}
    = \delta(\vr, \vr') \int\rd E  \sum_\nu g_\nu(E) {\big|\psi^{p}_{E\nu}(\vr)\big|}^2 
	  \frac{\partial f_\mathrm{FD}}{\partial E}(E - \mu_\nu)\,.
    \label{e:density_jacobian}
\end{equation}
Inserting the semi-classical approximation of the Jacobian in Eq.~(\ref{e:newton_poisson}) and rearranging yields a linear Poisson equation for each iteration $p+1$ of the Newton-Rhapson procedure:
\begin{equation}
	\Big( \mathrm{D} - \mathrm{M} \mathrm{J}^p_\rho \Big) \vec{V}^{p+1}
	= \mathrm{M} \Big( \boldsymbol{\rho}_\mathrm{net}\left[\vec{V}^p\right] - \mathrm{J}^p_\rho \vec{V}^p \Big) \,.
\end{equation}
The linear Poisson equation is a simple elliptic partial differential equation which is efficiently and accurately solved using the algebraic multi-grid (AMG) method.~\cite{Bell2011}
Figure.~\ref{f:sc_convergence} shows the convergence behaviour of a self-consistent calculation starting from a flat potential.
After an initial period, our Newton method, with a semi-classical approximation of the Jacobian, converges linearly.

To further accelerate the convergence of the self-consistent procedure, we use the Direct Inversion of the Iterative Subspace (DIIS) technique, commonly known as Pulay mixing in computational chemistry.~\cite{Pulay1982}
In the DIIS technique, the residual $\mathrm{R}^{p+1} = \vec{V}^{p+1} - \vec{V}^{p}$ and the new solution $\vec{V}^{p+1}$ are added to the previous solutions and form the iterative subspace, which is used to predict a new vector $\vec{\tilde{V}}^{p+1}$.
Following the analysis in Ref~\cite{Shepard2007}, we have implemented the DIIS technique using a least-squares approach based on the Singular-Value Decomposition (SVD) that improves the resolution of components of the iterative subspace when the residuals are almost linearly dependent.
This linear dependence occurs naturally when self-consistency is almost reached, and the tolerance for convergence is set very low.
However, to avoid a spurious linear-dependence that could degrade the convergence behavior, we limit the range of the iterative subspace.
Typically, only the last $5$ iterations are kept.
In practice, both the bare Newton and the accelerated DIIS method exhibit a linear convergence when solving the non-linear Poisson equation.
However, as demonstrated in Fig.~\ref{f:sc_convergence}, the DIIS method accelerates convergence by a factor of two, which is well worth the additional complexity of implementation.

\begin{figure}[htbp]
    \centering
    \includegraphics{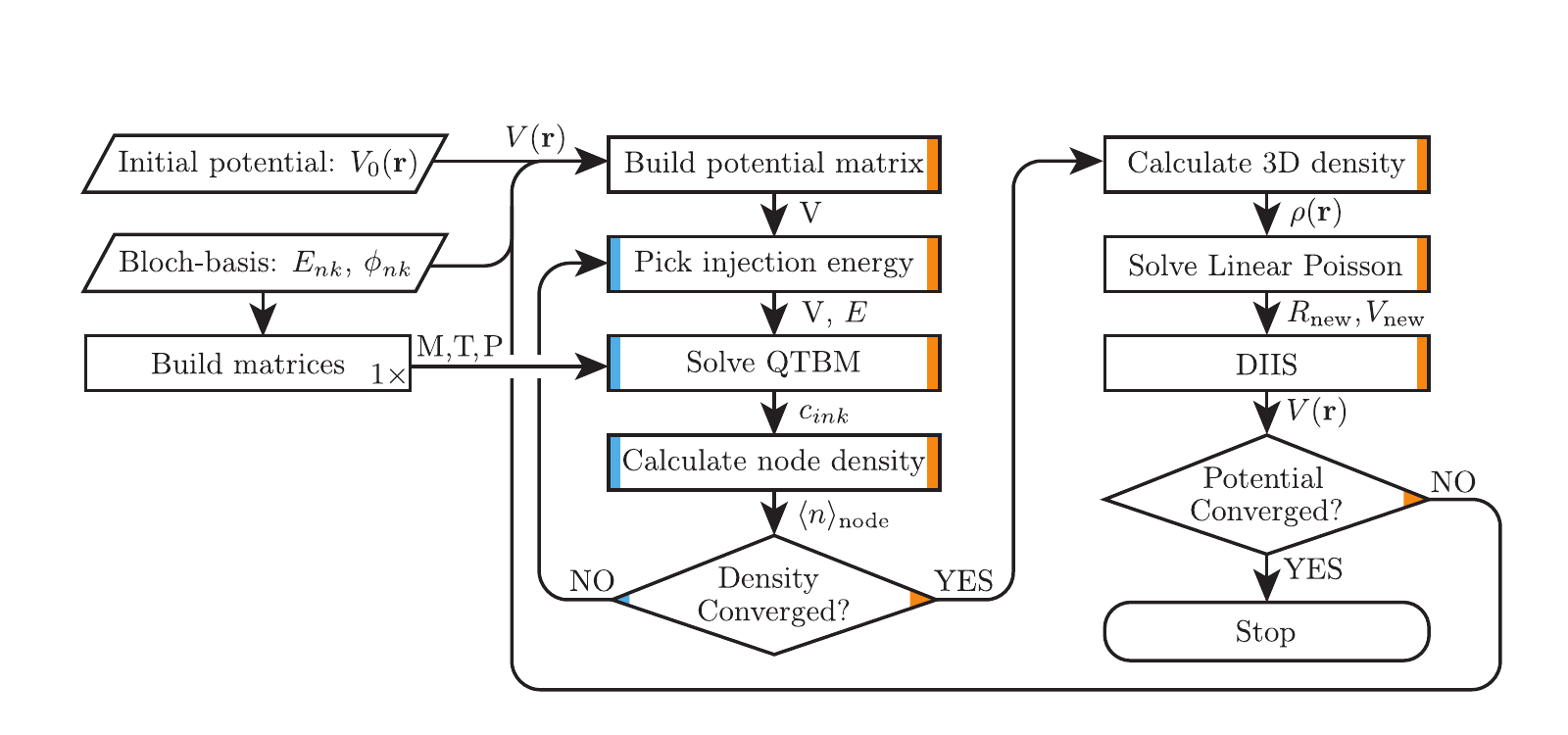}
    \caption{%
        Flowchart of the self-consistent procedure explained in the text.
		The inputs are an estimated initial potential and the Bloch-basis calculated using empirical pseudopotentials.
        The loop responsible for the adaptive Simpson-integration is indicated with blue shading on the left side, while the self-consistent loop is indicated with orange shading on the right hand side of the box.
        Note that the M, T, and P matrices are built a single time ($1\times$).
    }%
    \label{f:flowchart}
\end{figure}

Figure~\ref{f:flowchart} gives an overview of the entire self-consistent procedure for a typical simulation. 
Upon convergence, other quantities, such as the electronic current, can be calculated using the converged potential.

\section{Results: Graphene Nanoribbon}%
\label{s:results}

\begin{figure}[t]
    \centering
    \includegraphics{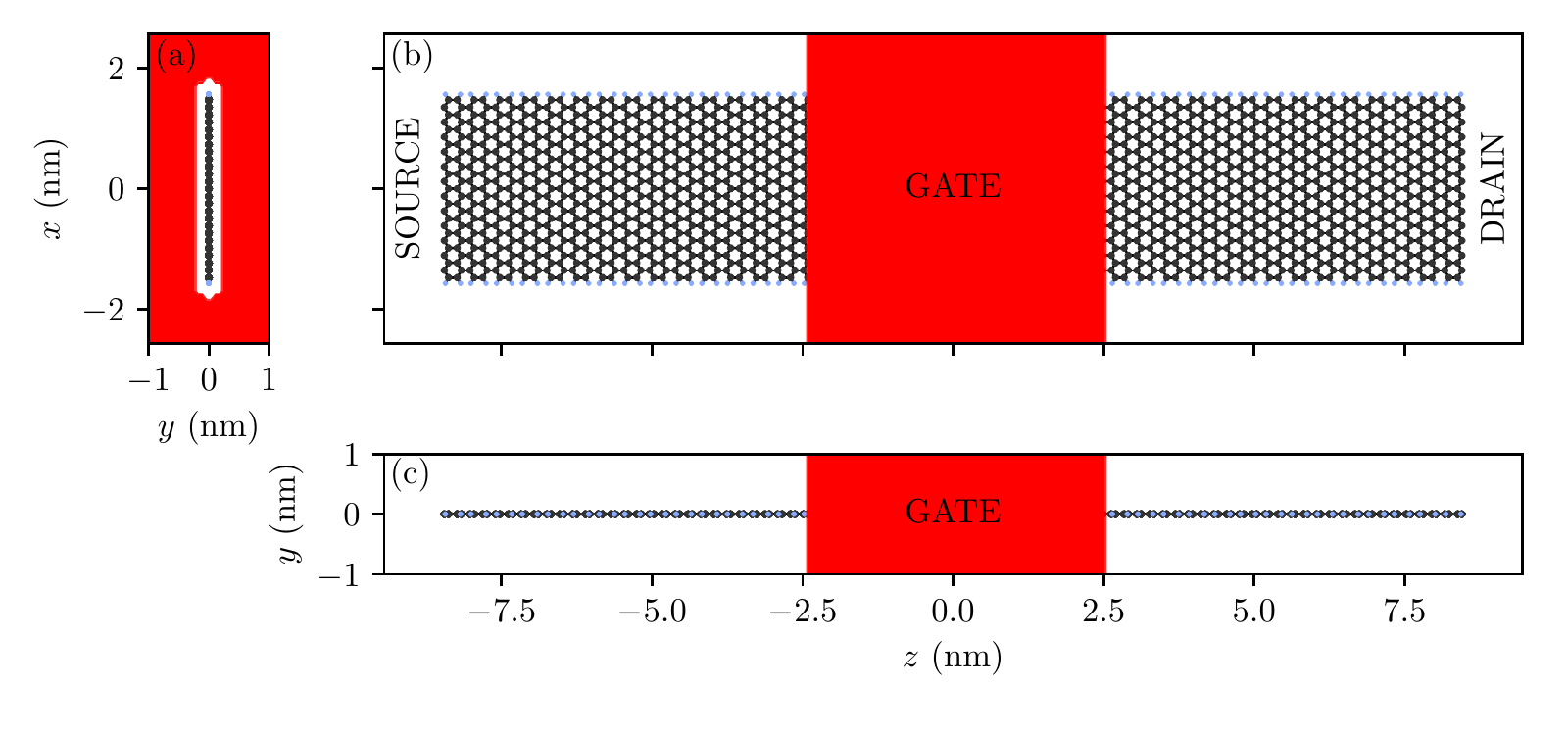}
    \caption{%
        (a) Front, (b) top, and (c) side view of a depiction of the aGNR FET under study.
        The armchair graphene-nanoribbon (aGNR) is $3.8\,\mathrm{nm}$ wide (25 carbon atoms).
        The simulated device is built from 40 repetitions of a single supercell, totalling 2160 atoms and a device length of approximately $17\,\mathrm{nm}$.
        The gate (shaded region) is centrally located in an all-around configuration, is $5\,\mathrm{nm}$ long, and has an oxide thickness equivalent to $1\,\mathrm{nm}$ of SiO$_2$.
        The source and drain terminals are assumed to be infinite extensions of the {aGNR}.
        The ribbon is uniformly n-type doped, except for the channel under the gate which is assumed to be p-type.
        Carbon (dark grey) and hydrogen (light blue) atom locations are indicated with spheres.
    }%
    \label{f:device}
\end{figure}

We demonstrate the Bloch wave method presented in Section~\ref{s:theory} using an armchair graphene nanoribbon (aGNR) field-effect transistor (FET), as shown in Fig~\ref{f:device}.
The aGNR is 25 carbon atoms wide (2 nm) and is terminated by hydrogen at the armchair edge.
The complete device is 17 nm long, with a channel length of 5 nm, and contains a total of 2160 atoms (2000 carbon and 160 hydrogen).

\subsection{Electronic Structure}

\begin{figure}[!htbp]
    \centering
    \includegraphics{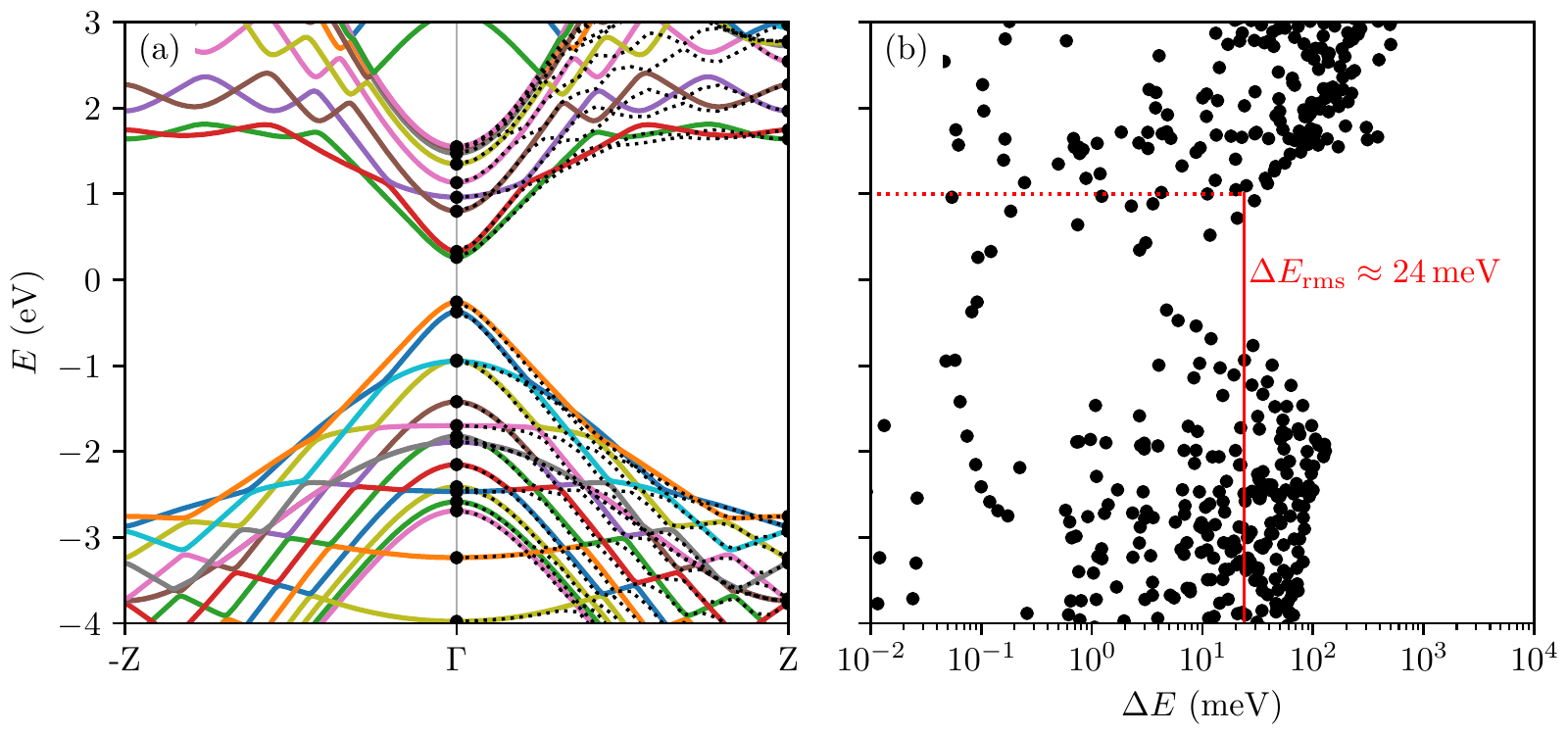}
    \caption{Reconstruction of the band structure using the Bloch-wave basis. (a) Band structure of the graphene nanoribbon shown in Fig.~(\ref{f:device}) near the Fermi level ($0$ eV), calculated using the plane-wave empirical pseudopotential method (dashed) and reconstructed from the Bloch waves at the $\Gamma$-point and  $\mathrm{Z}$-point (dots) (solid lines).
    (b) The absolute energy difference, $\Delta E$, between the full plane-wave method and the Bloch wave reconstruction. The root mean squared value of the energy difference, $\Delta E_\mathrm{rms}$, up to $1\,\mathrm{eV}$ is indicated on the graph.}%
    \label{f:bandstructure_reconstruction}
\end{figure}

We calculate the band structure of the $2\,\mathrm{nm}$ wide armchair GNR shown in Fig.~\ref{f:device} and show the results in Fig.~\ref{f:bandstructure_reconstruction}.
We first calculate the electronic structure using the plane-wave empirical pseudopotential method.
We use the local pseudopotentials from Ref.~\cite{Kurokawa2000} for both the carbon and hydrogen ions. 
Local pseudopotentials have been used extensively for carbon compounds and have been shown to accurately reproduce the band structure of graphene as well as its nanoribbons~\cite{Fischetti2011,Fischetti2016,Fang2017,Fang2016a,VandePut2016,Kurokawa2000}.
The resulting Schr\"odinger equation is solved in a plane-wave basis, using the fast Fourier transform for efficient evaluation, as described in Ref.~\cite{VandePut2016}.
In particular, we calculate both the eigenenergies, $\epsilon_{nk}$, and wavefunctions, \ie, the Bloch waves $\phi_{nk}(\vr)$, for 20 wave vectors, equally spaced from the first Brillouin zone (BZ) center ($\Gamma$-point) to its edge ($\mathrm{Z}$-point).
The resulting band structure is shown in Fig.~\ref{f:bandstructure_reconstruction} (a) as dotted lines.

To verify if this basis is capable of describing the electronic structure of the ribbon, we use the Bloch-wave expansion to reconstruct the band structure throughout the entire first {BZ}.
The Bloch waves, $\phi_{nk}(\vr)$, at the $\Gamma$-point and Z-point are used as a basis in our Bloch wave expansion of the wavefunction in Eq.~(\ref{e:wavef_expansion}).
The procedure is a straightforward application of Bloch's theorem: For a given wave vector $k'$, we calculate the expansion coefficients $c_{ink}$ at a single node $i$ by enforcing periodicity to the neighboring nodes with a phase-difference given by the wave vector $k'$,
\[
    c_{ink} = c_{jnk} \ee^{\ii k' (z_i - z_j)}\,.
\]
Exploiting this periodicity reduces the matrix equation, given in Eq.~(\ref{e:expansion_matrix_eq}), to a generalized eigenvalue problem of size ($N_\mathrm{Bloch} \times N_\mathrm{Bloch}$),
\[
    \mathrm{H}(k') \vec{c} = E_{k'} \mathrm{M}(k') \vec{c}\,,
\]
with:
\[
    \mathrm{H}(k') = H_{i,i} + \sum_{\langle i, j\rangle} H_{i,j}\, \ee^{\ii k' (z_i - z_j)}
    \quad\text{and}\quad
    \mathrm{M}(k') = M_{i,i} + \sum_{\langle i, j\rangle} M_{i,j}\, \ee^{\ii k' (z_i - z_j)}\,,
\]
for any node $i$ with nearest-neighbors $j$. Note that for the band structure, no external potential is applied and all external potential matrix elements vanish, \ie, $V_{ink} = 0$.

Figure~\ref{f:bandstructure_reconstruction} (a) shows the Bloch waves selected as the basis-set, and the band structure reconstructed using the Bloch wave method as solid lines.
In effect, we interpolate the band structure from the $\Gamma$-point and $Z$-point to the full first {BZ}.
Comparing the reconstructed band structure to the plane-wave calculation shows a good match.
Figure~\ref{f:bandstructure_reconstruction} (b) quantifies the error in the reconstruction, showing the absolute energy difference, $\Delta E$, between the Bloch-wave reconstructed band structure and the plane-wave values at the wave vectors of the plane-wave calculation.
As expected, the error increases in the upper conduction bands, where it is more likely that our selected Bloch-wave basis-functions do not span the solution space for every wave vector.
However, up to $1\,\mathrm{eV}$ above the Fermi level the Bloch-wave reconstruction matches the plane-wave results very well, showing a root mean squared error of $24\,\mathrm{meV}$.
This energy range is more than adequate for transport purposes. 
Moreover, if more accuracy is needed at a higher energy, one can readily increase the basis to cover those higher energies, albeit at an increased computational cost.

\begin{figure}[htbp]
    \centering
    \includegraphics{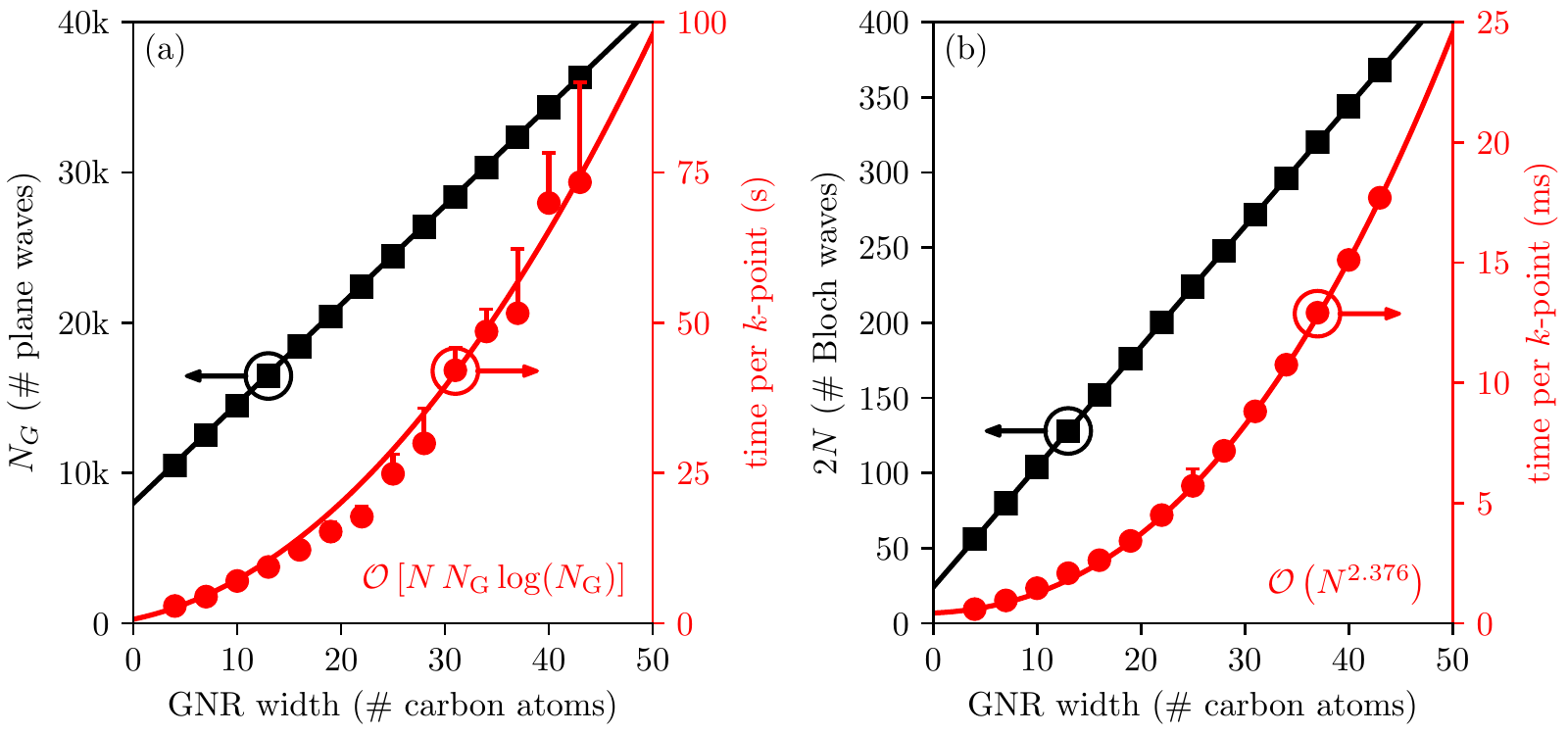}
    \caption{Computational time for the calculation of the band structure for various ribbon widths.
        (a) Using the plane-wave empirical pseudopotential method, with a basis of $N_G$ plane-waves.
        (b) Using our Bloch wave method, with $N$ Bloch waves taken at the first Brillouin-zone center and its edge.
        Note the different scales used in (a) and (b).
        The best computational time, out of seven runs, is indicated with a dot, while the range of the timing is shown with a bar.
        The ideal scaling behavior for each case is indicated and a fit is shown as a continuous curve. The basis set for the plane-wave method in (a) is 100 times larger than the Bloch wave basis used in (b). The timing shows an even greater speed-up than expected from the basis-set size alone.}%
    \label{f:timing_bandstructure}
\end{figure}

Having verified the Bloch-wave method's accuracy, we verify our earlier computational claims.
Figure~\ref{f:timing_bandstructure} shows the time required to calculate the eigenvalues for a single wave vector using (a) the plane-wave method and (b) the Bloch-wave reconstruction for different ribbon widths, as indicated by the number of carbon atoms from one edge to the other. 
Figure~\ref{f:timing_bandstructure} also shows the size of the basis set used for both methods.
The basis size corresponds to the number of plane-waves, $N_G$, for the plane-wave empirical pseudopotential method (a), and the number of Bloch waves $N_\mathrm{Bloch} = 2 * N$ for the Bloch-wave method (b), where $N$ is the number of bands.
In both methods, the basis set increases linearly with the ribbon size, scaling with the supercell length in (a) and scaling with the number of atoms (valence electrons) in (b).
However, for the range of GNR-widths shown here, the Bloch-wave basis-set is 100 times smaller than the plane-wave basis.
The most immediate effect of this reduction of basis-set size is a hundred-fold reduction in the required memory to store the coefficients $c_{ink}$ instead of all the plane-wave components of the wavefunction.
Therefore, we are able to avoid the single most limiting factor for the scaling of the envelope function approach to plane-wave based electron transport calculations~\cite{Fang2016,Fang2017}.
Note that, using the expansion in Eq.~(\ref{e:wavef_expansion}), we can obtain the real-space representation of the calculated coefficients $c_{ink}$ when needed, as we demonstrate in Section~\ref{ss:transport}.

The reduction of the size of the basis set also translates directly to a decrease of computing time.
For example, the band-structure calculations for the $2\,\mathrm{nm}$-wide 25-aGNR band-structure, shown in Fig.~\ref{f:bandstructure_reconstruction}, take $25$ seconds using the plane-wave method and only $5$ milliseconds using the Bloch wave method.
While this speedup of a factor of 5000 does not include the construction of the various overlap matrices required by the Bloch-method, these are pre-calculated only once.
Therefore, we expect equivalent performance gains for transport simulations.
Finally, comparing the computation time for different widths in Fig.~\ref{f:timing_bandstructure}, both methods show the scaling behaviour expected from their computational complexity.
The plane-wave method is bounded by the $\mathcal{O}(N_G \log N_G)$ complexity of the FFT algorithm, while the Bloch-wave method behaves in line with the $\mathcal{O}(N^{\approx 2.376})$ complexity of the matrix products, as implemented in the optimized Basic Linear Algebra Subprograms (BLAS)~\cite{Goto2008}.

\subsection{Transport: aGNR FET}%
\label{ss:transport}

The electron transport through the aGNR FET, shown in Fig.~\ref{f:device}, is calculated using the self-consistent procedure described in Section~\ref{s:theory}.
For our purposes, we apply a $0.2\,\mathrm{V}$ bias between source and drain, $V_\mathrm{ds}$.
We then vary the gate potential, $V_\mathrm{g}$, from $-0.7\,\mathrm{V}$ to $0.3\,\mathrm{V}$, calculate the potential self-consistently, and obtain the current through the device.
The work-function of the gate is set to the electron-affinity of the {aGNR}.

\begin{figure}[htbp]
    \centering
    \includegraphics{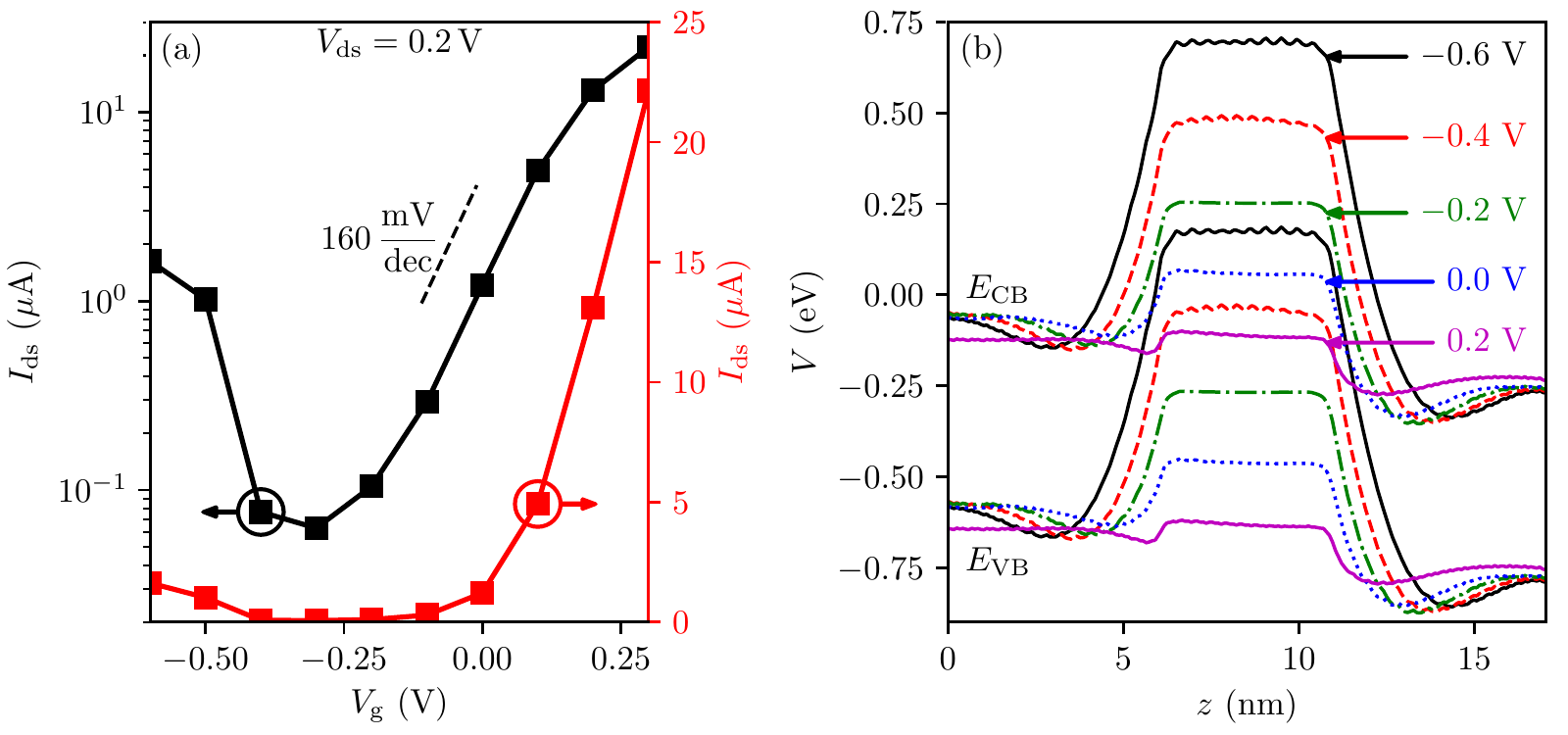}
    \caption{%
        Simulation results for device in Fig.~\ref{f:device}. 
        (a) Transfer-characteristics, showing source-drain current $I_\mathrm{ds}$ on a logarithmic (left) and linear (right) scale for different gate potentials $V_\mathrm{g}$.
    (b) Band-edge profile along $z$, through the middle of the ribbon, showing the approximate position of the conduction band minimum $E_\mathrm{CB}$ and valence band maximum $E_\mathrm{VB}$, for different gate potentials, as indicated.}%
    \label{f:iv_bandprofile}
\end{figure}

Figure~\ref{f:iv_bandprofile} (a) shows the obtained transfer-characteristics of the device.
Figure.~\ref{f:iv_bandprofile} (b) shows the corresponding band-profiles for different gate biasses, obtained self-consistently.
Under forward bias ($V_\mathrm{g} > 0$), the device operates as a conventional {FET}.
The sub-threshold and linear regimes are clearly visible in the figure.
The sub-threshold slope is about $160\,\mathrm{mV/dec}$. 
As already described for smaller ribbons~\cite{Fang2017}, this poor slope is caused by source-to-drain tunneling through the barrier in the channel, induced by the gate. 
These tunneling rates grow as the bandgap becomes smaller as the width of the ribbon increases.

For the simulated ribbon, the bandgap is only $0.52\ \mathrm{eV}$.
This small bandgap leads to interesting ambipolar behavior: the current increases if the gate is operated in reverse bias ($V_\mathrm{g} < 0$) due to band-to-band tunneling.
Looking at the band-alignment for, \eg, $V_\mathrm{g} = -0.6\ \mathrm{V}$, in Fig.~\ref{f:iv_bandprofile} (b), it is clear that band-to-band tunneling is possible from the source to the channel region, and once more towards the drain.
Thanks to the blocking of carriers from the high energy tail of the injected Fermi-Dirac distribution, the tunneling current increases at a steeper slope than in the forward regime.
This operating principle leading to the steep slope is the same as that of a Tunnel FET (TFET).~\cite{Verreck2015,Balaji2018}

Note that, in this device, the behavior of in the normal mode of operation as well as the reverse biased gate operation is based on quantum mechanics.
Our proposed method, offering an efficient full-band quantum mechanical transport solver for general atomistic structures, is naturally capable of dealing with these effects and provides an invaluable tool in the study of exotic materials and devices.

\section{Conclusions}%
\label{s:conclusions}

We have presented a numerical method for the atomistic calculation of quantum electron transport in nanoscaled structures using empirical pseudopotentials.
Our method is highly efficient; we have shown a reduction of the size of the required computational basis by two orders of magnitude compared to the conventional plane-wave methods.
This efficiency is achieved by treating differently the two length-scales in the system. 
A partition-of-unity captures the large-scale behaviour of the system and admits only nearest-neighbor coupling, resulting in excellent scalability.
The atomic scale, meanwhile, is captured by an expansion based on Bloch waves of the atomic structure at high symmetry points.
The Bloch-waves are computed to high accuracy using a Fourier-based plane-wave approach before starting the transport calculations.
Our method approximates the computational efficiency of tight-binding and mode-space approaches while retaining the advantages of the plane-wave method, which features a natural real space representation with sub-atomic resolution.

We solve the electronic states in our open system using the well known quantum-transmitting boundary method (QTBM) and update self-consistently the Hartree potential from the three-dimensional density.
The density is obtained by adaptively integrating the individual wave-functions' densities.
Notably, we systematically control and estimate the numerical error at each stage in our method by using iterative solvers and adaptive integration methods.
We are thus assured that the accuracy of our results is limited by the physical approximations made, and not by the numerical errors.

We have demonstrated the accuracy and efficiency of our method by calculating the ballistic transport properties of a graphene nanoribbon transistor.
In this test case, the reconstruction of the band structure from our Bloch-wave basis is accurate to $24\,\mathrm{meV}$ when compared to the full-plane wave calculation, while being three orders of magnitude faster more efficient.
Comparing different widths of nanoribbon shows that our method scales as expected, with a significantly improved performance compared to previous plane-wave approaches.
A hundred-fold reduction in the size of the basis set results in a similar reduction in the memory requirements, which severely limit the device-size that can be handled by previous plane-wave envelope-function approaches.
As a demonstration, we have simulated transport in a $3\,\mathrm{nm}$ wide nanoribbon transistor.
We have observed a significant deterioration of the sub-threshold behavior due to source-to-drain tunneling through the potential barrier induced by the gate bias.
In reverse bias, we observe significant ambipolar current due to band-to-band tunneling through the small bandgap.
This reaffirms the need for a quantum mechanical treatment of the transport in nanostructured devices in the `intermediate' nanoscale, between bulk crystalline behavior and few-atom devices.
Our presented method provides an efficient and flexible basis for such studies.

Finally, while we have only illustrated our method using empirical pseudopotentials, our approach is generally applicable to any formulation that can provide the Bloch waves in a real-space basis.
Of particular interest might be the various \textit{ab-initio} methods based on plane-waves, for which electron transport calculations are prohibitively expensive.

\section*{Acknowledgements}

This material is based in part on work supported by the National Science Foundation under Award Number 1710066.
Any opinions, findings, and conclusions or recommendations expressed in this material are those of the authors and do not necessarily reflect the views of the National Science Foundation.

\appendix

\section{Matrix element of the crystal Hamiltonian}%
\label{a:weak_bloch}

We derive an expression for the matrix elements for the crystal Hamiltonian that avoids explicit knowledge of the crystal potential,
\begin{equation}
    \mathrm{H}^\mathrm{c}_{i'n'k',ink} 
    = \intall f^*_{i'}(\vr) \phi^*_{i'n'k'}(\vr) 
              \left[-\frac{\hbar^2}{2m}\vnabla^2 + V^\mathrm{c}_i(\vr)\right]
              \big[ f_i(\vr) \phi_{ink}(\vr) \big]\,.
    \label{e:hc_matel}
\end{equation}
To remove the crystal potential, we start from the known Schr\"odinger equation for the Bloch waves in a supercell
\begin{equation}
    \left[ -\frac{\hbar^2}{2m} \vnabla^2 + V^\mathrm{c}_i(\vr)\right]  \phi_{ink}(\vr)
     = \epsilon_{ink}\, \phi_{ink}(\vr) \,.
\end{equation}
We left-multiply by $f^*_{i'}(\vr) \phi^*_{i'n'k'}(\vr) f_{i}(\vr)$ and integrate over all of space, yielding
\begin{equation}
    \intall
      f^*_{i'}(\vr) \phi^*_{i'n'k'}(\vr) f_{i}(\vr)
      \left[-\frac{\hbar^2}{2m}\vnabla^2 + V^\mathrm{c}_i(\vr)\right]
      \phi_{ink}(\vr) 
    = \mathrm{M}_{i'n'k',ink}\, \epsilon_{ink} \,,
    \label{e:bloch_se}
\end{equation}
where we have defined the overlap matrix element as
\begin{equation}
    \mathrm{M}_{i'n'k',ink} 
    = \intall f^*_{i'}(\vr) \phi^*_{i'n'k'}(\vr) 
               f_i(\vr) \phi_{ink}(\vr)\,.
\end{equation}

Comparing Eq.~(\ref{e:bloch_se}) to the matrix element of the crystal Hamiltonian in Eq.~(\ref{e:hc_matel}) we obtain
\begin{equation}
    \mathrm{H}^\mathrm{c}_{i'n'k',ink} 
    = \mathrm{M}_{i'n'k',ink}\, \epsilon_{ink}
    + \mathrm{T}^\mathrm{(r)}_{i'n'k',ink}
    + \mathrm{P}^\mathrm{(r)}_{i'n'k',ink}\,,
    \label{e:hc_matel_r}
\end{equation}
where we have defined additional matrix elements representing kinetic energy and momentum-coupling
\begin{align}
    \mathrm{T}^\mathrm{(r)}_{i'n'k',ink} 
    &= -\frac{\hbar^2}{2m}
       \intall f^*_{i'}(\vr) \phi^*_{i'n'k'}(\vr) 
               \big[ \vnabla^2 f_i(\vr) \big] 
               \phi_{ink}(\vr)\,, \\
    \mathrm{P}^\mathrm{(r)}_{i'n'k',ink} 
    &= -\frac{\hbar^2}{m}
       \intall f^*_{i'}(\vr) \phi^*_{i'n'k'}(\vr) 
               \big[ \vnabla f_i(\vr) \big] \cdot
               \big[ \vnabla \phi_{ink}(\vr) \big] \,,
\end{align}
where the subscript (r) is a reminder that the matrix elements are non-Hermitian and the operators they contain act only to the right.
Similarly, we can define ``left'' matrix elements
\begin{align}
    \mathrm{T}^\mathrm{(l)}_{i'n'k',ink} 
    = {\left(\mathrm{T}^\mathrm{(r)}_{ink,i'n'k'} \right)}^* 
    &= -\frac{\hbar^2}{2m}
       \intall \big[ \vnabla^2 f^*_{i'}(\vr) \big]
               \phi^*_{i'n'k'}(\vr) 
               f_i(\vr) \phi_{ink}(\vr)\,, \\
    \mathrm{P}^\mathrm{(l)}_{i'n'k',ink} 
    = {\left(\mathrm{P}^\mathrm{(r)}_{ink,i'n'k'} \right)}^*
    &= -\frac{\hbar^2}{m}
       \intall \big[ \vnabla f^*_{i'}(\vr) \big] \cdot
               \big[ \vnabla \phi^*_{i'n'k'}(\vr) \big]
               f_i(\vr) \phi_{ink}(\vr)\,,
\end{align}
that satisfy
\begin{equation}
    \mathrm{H}^\mathrm{c}_{i'n'k',ink} 
    = \epsilon_{i'n'k'}\, \mathrm{M}_{i'n'k',ink}
    + \mathrm{T}^\mathrm{(l)}_{i'n'k',ink}
    + \mathrm{P}^\mathrm{(l)}_{i'n'k',ink}\,,
    \label{e:hc_matel_l}
\end{equation}
Combining the expressions of Eq.~(\ref{e:hc_matel_r}) and Eq.~(\ref{e:hc_matel_l}) yields
\begin{equation}
    \mathrm{H}^\mathrm{c}_{i'n'k',ink} 
    = \frac{\epsilon_{ink} + \epsilon_{i'n'k'}}{2}\, \mathrm{M}_{i'n'k',ink}
    + \mathrm{T}_{i'n'k',ink}
    + \mathrm{P}_{i'n'k',ink}\,,
    \label{e:hc_matel_herm}
\end{equation}
in which
$\mathrm{T}_{i'n'k',ink} = \left(\mathrm{T}^\mathrm{(r)}_{i'n'k',ink} +  \mathrm{T}^\mathrm{(l)}_{i'n'k',ink} \right) /2$ and
$\mathrm{P}_{i'n'k',ink} = \left(\mathrm{P}^\mathrm{(r)}_{i'n'k',ink} +  \mathrm{P}^\mathrm{(l)}_{i'n'k',ink} \right) /2$
are both Hermitian matrix elements.

For the correct preservation of probability current across nodes, and in particular if we intend to use linear shape functions for $f_i(\vr)$, we should further use integration by parts in the derivation of the kinetic matrix elements,
\begin{multline}
    \mathrm{T}^\mathrm{(r)}_{i'n'k',ink} 
    = \frac{\hbar^2}{2m}
      \intall \Big\{ 
          \big[ \vnabla f^*_{i'}(\vr) \big]
          \phi^*_{i'n'k'}(\vr) \cdot
          \big[ \vnabla f_i(\vr) \big] 
          \phi_{ink}(\vr)
        + f^*_{i'}(\vr)
          \big[ \vnabla \phi^*_{i'n'k'}(\vr) \big]\cdot
          \big[ \vnabla f_i(\vr) \big] 
          \phi_{ink}(\vr)\\
        + f^*_{i'}(\vr)
          \phi^*_{i'n'k'}(\vr) 
          \big[ \vnabla f_i(\vr) \big]\cdot 
          \big[ \vnabla \phi_{ink}(\vr) \big]
      \Big\}\,,
\end{multline}
where we have omitted the vanishing boundary term.
Combining this result with its Hermitian conjugate, and grouping appropriate terms, yields a compact form for the Hermitian kinetic energy matrix elements,
\begin{equation}
    \mathrm{T}_{i'n'k',ink}
    = \frac{\hbar^2}{2m}
      \intall \Big\{ \frac{1}{2}
          \vnabla \big[ f^*_{i'}(\vr) f_i(\vr) \big]
          \cdot
          \vnabla \big[ \phi^*_{i'n'k'}(\vr) \phi_{ink}(\vr) \big]
      + \big[ \vnabla f^*_{i'}(\vr) \big]
        \phi^*_{i'n'k'}(\vr) \cdot
        \big[ \vnabla f_i(\vr) \big] 
        \phi_{ink}(\vr)
     \Big\}\,.
\end{equation}

\bibliographystyle{elsarticle-num}
\bibliography{\jobname}

\end{document}